\documentclass[twocolumn,trackchanges]{aastex62}
 \usepackage{graphicx}
 \usepackage{natbib}
 \usepackage{times}
 \usepackage{amsmath}
 \usepackage{textcomp}
 \usepackage{longtable}
 \usepackage{xcolor}
 \usepackage{booktabs}
 \usepackage[colorinlistoftodos]{todonotes}
 \usepackage{lineno}
 \usepackage{balance}

\newcommand{\Msun}{\ensuremath{M_{\odot}} }

\submitjournal{ApJ}
\shorttitle{Hunting distant BL Lacs with the photometric technique using \emph {Swift} and SARA}
\shortauthors{Rajagopal et al.}

\def\asec{\ifmmode^{\prime\prime}\else$^{\prime\prime}$\fi}
\def\degs{\ifmmode ^{\circ}\else$^{\circ}$\fi}

\begin{document}

\title{Hunting distant BL Lacs with the photometric technique using \emph {Swift} and SARA}
\email{changar@g.clemson.edu}
\author[0000-0002-8979-5254]{ M. Rajagopal}
\affil{Department of Physics and Astronomy, Clemson University, SC 29634-0978, U.S.A.\footnote{Astronomer at SARA Observatory}}
\author{ A. Kaur} 
\affil{Department of Astronomy and Astrophysics, 525 Davey Lab, Pennsylvania State University, University Park, 16802, USA}
\author{ M. Ajello}
\affil{Department of Physics and Astronomy, Clemson University, SC 29634-0978, U.S.A.}
\author{ A. Rau}
\affil{Max-Planck-Institut f\"ur extraterrestrische Physik, Giessenbachstra{\ss}e 1, 85748 Garching, Germany}
\author{ A. Dom\'{i}nguez}
\affil{IPARCOS and Department of EMFTEL, Universidad Complutense de Madrid, E-28040 Madrid, Spain}
\author{ B. Cenko} 
\affil{Astrophysics Science Division, NASA Goddard Space Flight Center, Mail Code 661, Greenbelt, MD 20771, USA}
\author{ J. Greiner}
\affil{Max-Planck-Institut f\"ur extraterrestrische Physik, Giessenbachstra{\ss}e 1, 85748 Garching, Germany}
\author{ D. H. Hartmann}
\affil{Department of Physics and Astronomy, Clemson University, SC 29634-0978, U.S.A.}
\affil{Southeastern Association for Research in Astronomy (SARA), USA}
 
\begin{abstract}
BL Lacertae objects (BL Lacs) represent a large fraction (22\%) of $\gamma$-ray sources in the Third {\it Fermi} Large Area Telescope catalog (3FGL). Nearly half of the BL Lac population remains without a redshift because of their featureless optical spectra. We aim to increase the number of BL Lacs with a redshift measurement by using the photometric technique. For this work, we have used 6 $\emph{Swift}$-UVOT filters and SDSS $g\arcmin, r\arcmin, i\arcmin, z\arcmin$ optical filters mounted on the 0.65 m SARA-CTIO located in Chile and the 1.0 m SARA-ORM in Canary Islands. A sample of 45 sources was selected from the 3FGL catalog for which photometry was performed in 10 optical and UV filters to obtain redshift measurements. We found 3 sources with $z>1.3$ while reliable upper limits have been provided for 17 sources. The results presented here bring the total number of high-$z$ ($z>1.3$) BL Lacs to 29. \\
\end{abstract}

\section{Introduction} \label{sec:intro}
The Third {\it Fermi} Large Area Telescope (LAT) source catalog (\citealp{3fgl}, 3FGL) detected over 1500 sources in the 100 MeV$-$300 GeV range, that belong to the blazar class. Blazars are an extreme class of active galaxies that have their relativistic jets aligned by a very small angle with our line of sight. Their spectral energy distribution (SED) displays two characteristic bumps, with one at lower energies (infrared to X-rays)  attributed to synchrotron emission, while the high-energy one (X-ray to $\gamma$-rays) is ascribed to synchrotron self-Compton emission \citep{maraschi1994} or inverse Compton scattering on external photon fields \citep{Sikora1994, ghisellini1996,  ghis1997, dermer1997}. \\
Blazars are also classified based on their optical spectra into flat spectrum radio quasars (FSRQs) and BL Lacertae objects (BL Lacs). FSRQs are known for the presence of broad emission lines (equivalent width $\textgreater$5\AA) in their spectra while BL Lacs have no or very weak emission lines \citep{Urry1995}. 
They dominate the 3FGL catalog with 484 FSRQs and 660 BL Lac objects.\\
\citet{padovani95} introduced another classification scheme of blazars based on the position of the synchrotron peak frequency, $\nu_{pk}^{sy}$. \citet{Abdo2010} subsequently identified the ranges of peak frequency for each category. They divided the blazar population into three different classes: Low Synchrotron Peaked blazars (LSP) where $\nu_{pk}^{sy}<10^{14}$ Hz; High Synchrotron Peaked blazars (HSP) with $\nu_{pk}^{sy}>10^{15}$ Hz and Intermediate Synchrotron Peaked blazars (ISP) for $10^{14}$ Hz $<\nu_{pk}^{sy}<10^{15}$ Hz. BL Lacs span the three classes (LSP, ISP and HSP; \citealp{3lac}) while FSRQs mainly belong to the LSP category. \\
Because of their hard GeV spectra, HSP BL Lacs are valuable in the study of extragalactic background light (EBL), which constitutes the integrated radiation from all stars, galaxies, and other objects in the universe since the re-ionization epoch \citep{Dominguez2015, fermi2018}.\\
Measuring the EBL directly is a challenge due to the presence of the zodiacal light as well as the emission from our Galaxy \citep{Hauser2001}. Therefore, an indirect method employed to measure the EBL makes use of distant $\gamma$-ray emitters. Photons from  $\gamma$-ray sources interact with the EBL photons and cause production of electron-positron pairs, which imprints a characteristic absorption signature in the spectra of these $\gamma$-ray sources \citep{Stecker1992,  Ackermann1190}. This attenuation in their spectra can be used to study the EBL evolution with redshift \citep{Aharonian2006} and other cosmological properties \citep{domprada, dom2019}. Furthermore, the higher the redshift of the source, the stronger will be the attenuation, hence providing a better EBL constraint. Thus, the requirement for  redshift measurements of $\gamma$-ray sources and particularly of HSP BL Lacs is imperative. \\
The photometric technique introduced for BL Lacs by \citet{Rau2012} proves to be well-suited for measuring the redshifts of BL Lacs that otherwise lack a spectroscopic redshift. The photometric method is based on the following principle: UV photons from a $\gamma$-ray source (BL Lac) are absorbed by the neutral Hydrogen along our line of sight causing a clear attenuation in the flux at the Lyman limit (912 \AA). The position of this dropout in the SED is modeled to measure the redshift of the BL Lac.\\ 
Applying this method to a sample of 103 BL Lacs (with no redshift measurements) from {\it Fermi's} Second Catalog of Active Galactic Nuclei (2LAC) \citep{2lac}, \citet{Rau2012} found 9 BL Lacs at high redshifts ($z\geq1.3$; out of which 6 were newly detected). Furthermore, \citet{Kaur2017} found 5 more BL Lacs from a sample of 40 sources and 2 more were found from a sample of 15 sources in \citet{Kaur2018}, bringing the number of confirmed high-$z$ BL Lacs to a total of 26 (19 being already reported in {\it Fermi's} 3LAC, \cite{3lac}, catalog). \\
In this work we employ the photometric technique to provide new BL Lac redshifts using a sample of 45 sources. We use a flat $\Lambda$CDM cosmological model with $H_{0}=70$ km s$^{-1}$ Mpc$^{-1}$, $\Omega_{m}=0.3$ and $\Omega_{\Lambda}=0.7$ for all calculations. All errors have been given at 1$\sigma$ confidence unless stated otherwise.\\
The organization of this paper is as follows: Section \ref{sec:obs} and \ref{sec:ana} describe the observations and data analysis, respectively. 
The details of the SED fitting technique are reported in Section \ref{sec:sed}, while the discovered high-$z$ BL Lacs are presented in Section \ref{sec:res}. Finally, Section \ref{sec:dis} summarizes our results.

\section{OBSERVATIONS} \label{sec:obs}
\subsection{Sample Selection and Observations}
Our sample includes 45 BL Lacs without a measured redshift selected from the 3FGL catalog. An approved cycle 13 program\footnote{Proposal: 1316180, PI: Dr. A. Kaur} with the Neil Gehrels $\emph{Swift}$ Observatory \citep{gehrels2004} allowed us to gather UVOT (UV/Optical Telescope; \citealp{Roming2005}) data for 8 sources. The rest of the sample has been observed by {\it Swift}  as Targets of Opportunity (ToO).\\
All the sources were observed with the SARA (Southeastern Association for Research in Astronomy) consortium's 0.65\,m and 1.0\,m telescopes located in Cerro Tololo, Chile (SARA-CT) and the Roque de los Muchachos Observatory in Canary Islands (SARA-ORM), respectively \citep{keel2017}. The data from the two aforementioned facilities were collected using SDSS $g\arcmin$, $r\arcmin$, $i\arcmin$, and $z\arcmin$ filters. Each source was observed with an exposure time ranging from 30-60 minutes (per filter). The sources were also observed with $\emph{Swift}$-UVOT in 6 UV-Optical filters ({\it uvw2, uvm2, uvw1, u, b, v}) for $\sim$2000 seconds. The combined data from the $\emph{Swift}$-UVOT and SARA telescopes provided us 10 filter flux measurements for each source. Details of the observations have been presented in Table \ref{tab:obs}. 

\section{Data Analysis}\label{sec:ana}

\subsection{SARA and Swift-UVOT}
Data from the SARA telescopes were obtained using the SDSS filters ($g\arcmin, r\arcmin, i\arcmin, z\arcmin$) and analyzed using the aperture photometry technique with IRAF v2.16 \citep{Tody1986}. Calibrations for the four optical filters were performed using standard star data in the SDSS Data Release 13 \citep{sdssdr13} or Landolt Equatorial Standards \citep{Landolt2009}. Correction for foreground Galactic extinction was performed using E(B-V) from \citet{schlafly2011}. \\
For $\emph{Swift}$-UVOT data reduction, we used the standard UVOT pipeline procedure provided by \citet{Poole2007}. This process removes bad pixels, flat-fields and corrects for the system response. The UVOT tasks, {\tt\string UVOTIMSUM} and {\tt\string UVOTSOURCE} using HEASoft (within HEASoft v.6.21\footnote{\url{https://heasarc.nasa.gov/lheasoft/}}) were employed to combine images from multiple observations and extract the magnitude of the sources, respectively. To this end, a circular region of radius (3.5$\arcsec$ - 5$\arcsec$) was used for source extraction for each object in order to maximize the signal-to-noise ratio. For background subtraction, an annulus of inner radius 8$\arcsec$ - 10$\arcsec$ and outer radius 20$\arcsec$ - 25$\arcsec$ was used for each source. The magnitudes obtained were then corrected for Galactic foreground extinction using Table 5 in \citet{kataoka2008}. 

\subsection{Cross-Calibration and Variability Correction} 
We follow a similar approach utilized by \citet{Rau2012} in order to calibrate between SARA and $\emph{Swift}$-UVOT filter bands. The spectral overlap between SDSS $g\arcmin$ filter (on SARA) and UVOT {\it b} filter is used to cross-calibrate the two instruments. This was achieved by assuming that the SED remains unchanged and can be approximated by a power law. We obtained theoretical magnitudes for all filters using the power law templates with spectral indices ($\beta$) between 0.3 and 3.0. The colors obtained by subtracting the magnitudes $g\arcmin-r\arcmin$ and {\it b}$\, -\, g\arcmin$ were plotted and fit with a quadratic curve. This provided us with the relationship reported in Eq.\ \ref{eqn:eq}. The offsets resulting from this equation in the {\it b} band were applied to the UVOT filter magnitudes. \\
\begin{equation}\label{eqn:eq}
b-g^\prime = 0.26\,(g^\prime-r^\prime)+0.02\,(g^\prime-r^\prime)^2  
\end{equation}
Furthermore, blazars are known to be highly variable throughout the electromagnetic spectrum on timescales that can vary from a few minutes to years. This variability can significantly contribute to the uncertainties of the SED constructed using multi-wavelength and non-simultaneous observations from multiple instruments.
Both SARA telescopes and $\emph{Swift}$-UVOT gather data sequentially in 4 optical and 6 UV-Optical filters, respectively. 
A systematic uncertainty of $\Delta m = 0.1$ mag was applied for each UVOT filter, following \citet{Rau2012}, to account for the variability between the exposures of each UVOT filter. We further include an uncertainty of $0.1$ mag for each SARA filter to account for the same. The corrected magnitudes for all the sources were converted to the AB system and have been reported in Table \ref{tab:mags}.

\begin{figure}[ht!]
\centering
\hspace{-0.7cm}
\includegraphics[width=1.05\columnwidth,trim=15 0 0 0,clip=true]{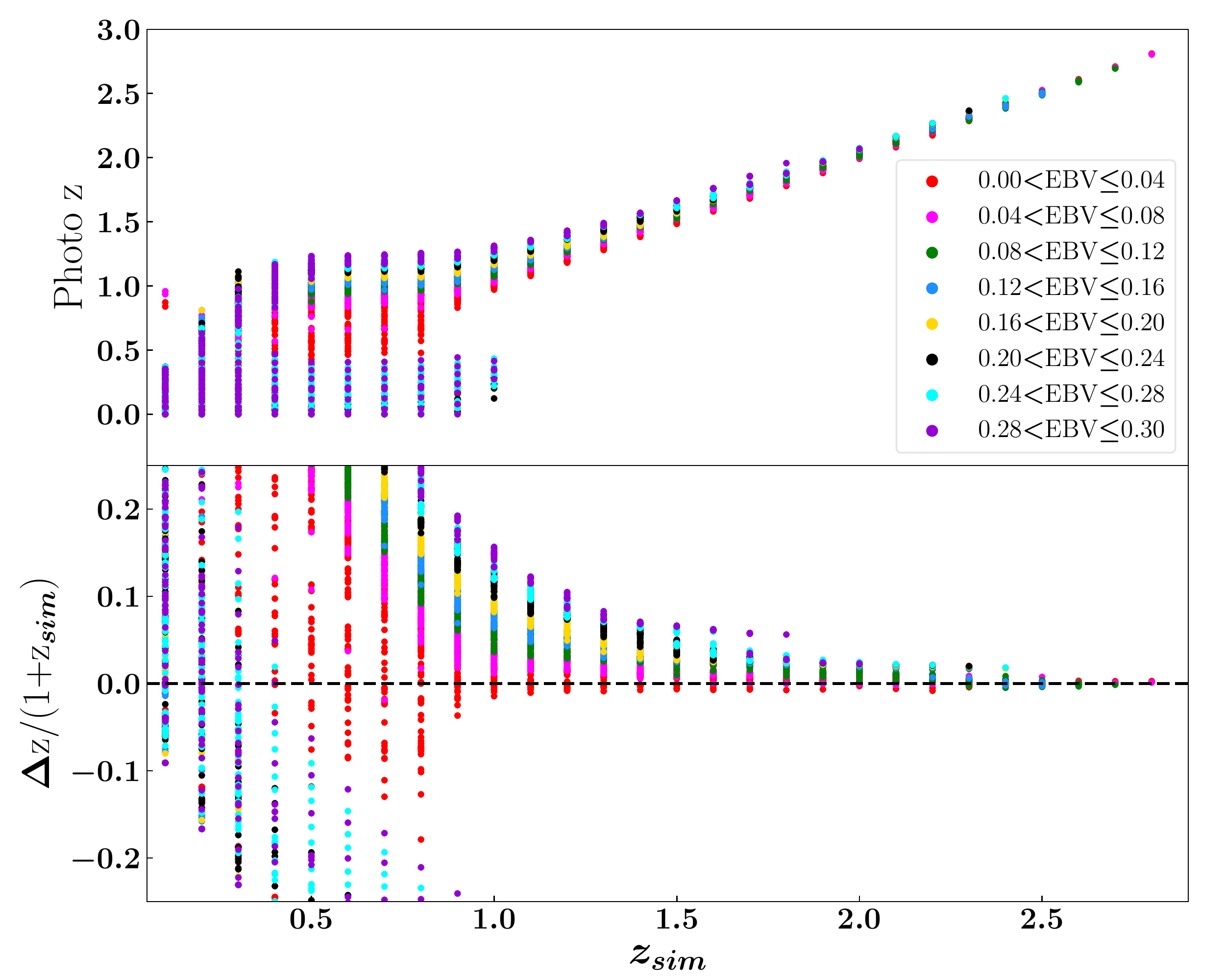}
\caption{Best-fit photometric redshifts vs. simulated redshifts for all the sources. Simulations for E(B-V)$\leq$0.30 ({\it top}) gives a minimum limit for our best-fit photo-$z$ value at $z_{best,phot}\sim1.3$ within an accuracy  $\left| \Delta z/(1+z_{sim}) \right| <0.015 $.}
\label{fig:sim2}
\end{figure}

\section{SED Fitting}\label{sec:sed}

We have employed the use of the publicly available software \texttt{LePhare v.2.2} \citep{Arnouts1999, Ilbert2006} for measuring photometric redshifts for our sources. This FORTRAN based code performs SED fitting by evaluating the differences between observational data and theoretical models based on the $\chi^{2}$ statistic. A set of libraries for stars, galaxies and quasars, provided in the \texttt{LePhare} package were used to fit the data. We employed 60 power-law templates of the form $F_{\lambda} \propto \lambda^{-\beta}$ where the value of $\beta$ was chosen to be in the range 0 to 3, in steps of 0.05. This was done under the assumption that the UV-Optical-Near-Infrared regime for BL Lacs can be approximated by a power-law model. A library of galaxies and galaxy/AGN hybrid templates \citep{Salvato2009, Salvato2011} was also used for the fitting procedure in order to distinguish between a low-redshift galaxy/AGN hybrid and a high-$z$ blazar in the photometric redshift solutions. The final library containing stellar templates was also employed for the fitting to check for any false associations \citep{Bohlin1995, Pickles1998, Chabrier2000}. \\
To assess the accuracy of the photometric redshift estimates, we performed simulations using the \texttt{LePhare} package to test power law SEDs with $\beta$ from 0 to 3, for redshifts ranging from $z=0$ to $z=3$ and for reddening, E(B-V), values up to 0.30. For each source, the input magnitudes were extracted from a Gaussian distribution of the model magnitudes taking into account statistical and systematic uncertainties arising from the calibration and variability. The resulting SEDs were fit using \texttt{LePhare} to derive the measured photometric redshifts ($z_{phot}$) with the input values.\\
As shown in Fig.~\ref{fig:sim2}, we found good agreement between the simulated and photo-$z$ values for $z>1.3$, provided reddening values, E(B-V)$\leq0.30$ within an accuracy of $\left| \Delta z/(1+z_{sim}) \right| <0.015 $. For all other sources, except those with E(B-V)$>0.3$, we can provide a reliable $z\lesssim1.3$ redshift upper limit.\\
The integral of the probability distribution function, $P_{z}$ (that describes the probability that the redshift of a source is within $0.1(1+z_{phot})$ of the best fit value) is another criterion employed to estimate the accuracy of this technique. For this work, results with $P_z > 90 \%$ were chosen as reliable photometric redshifts.\\

\section{Results}\label{sec:res}
\begin{figure}[ht!]
\centering
\hspace{-0.6cm}
\includegraphics[width=1.05\columnwidth,trim=10 0 0 0,clip=true]{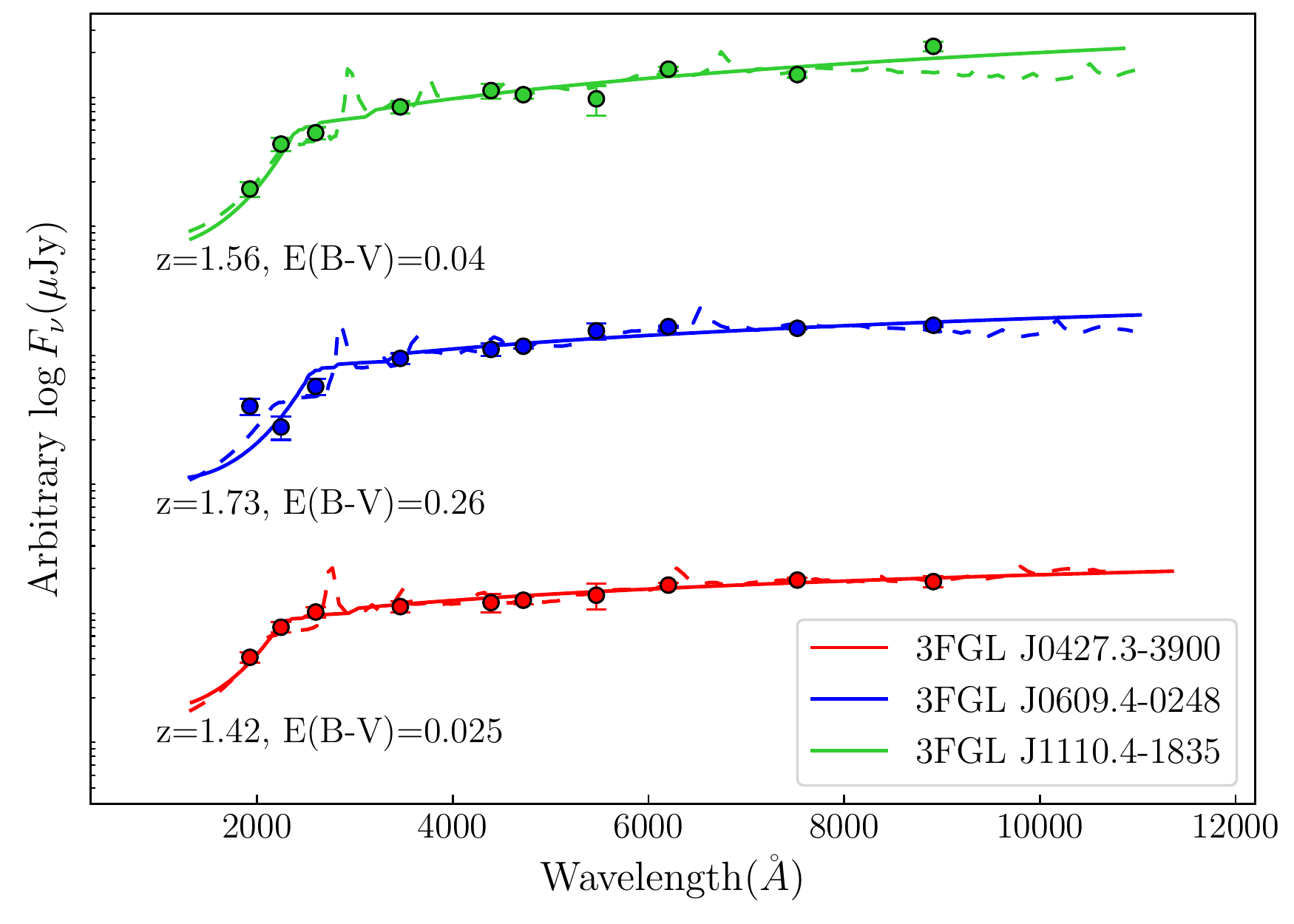}
\caption{\label{fig:sed}The $\emph{Swift}$-UVOT + SARA SED of the 3 high-$z$ sources. The solid line shows the power law template fits for each source with a clearly seen dropout towards the shorter wavelengths. In addition, the dashed line shows the best-fitting galaxy template to these objects.} 
\end{figure}

\begin{figure}[ht!]
\centering
\hspace{-0.6cm}
\includegraphics[width=1.05\columnwidth,trim=0 0 0 17,clip=true]{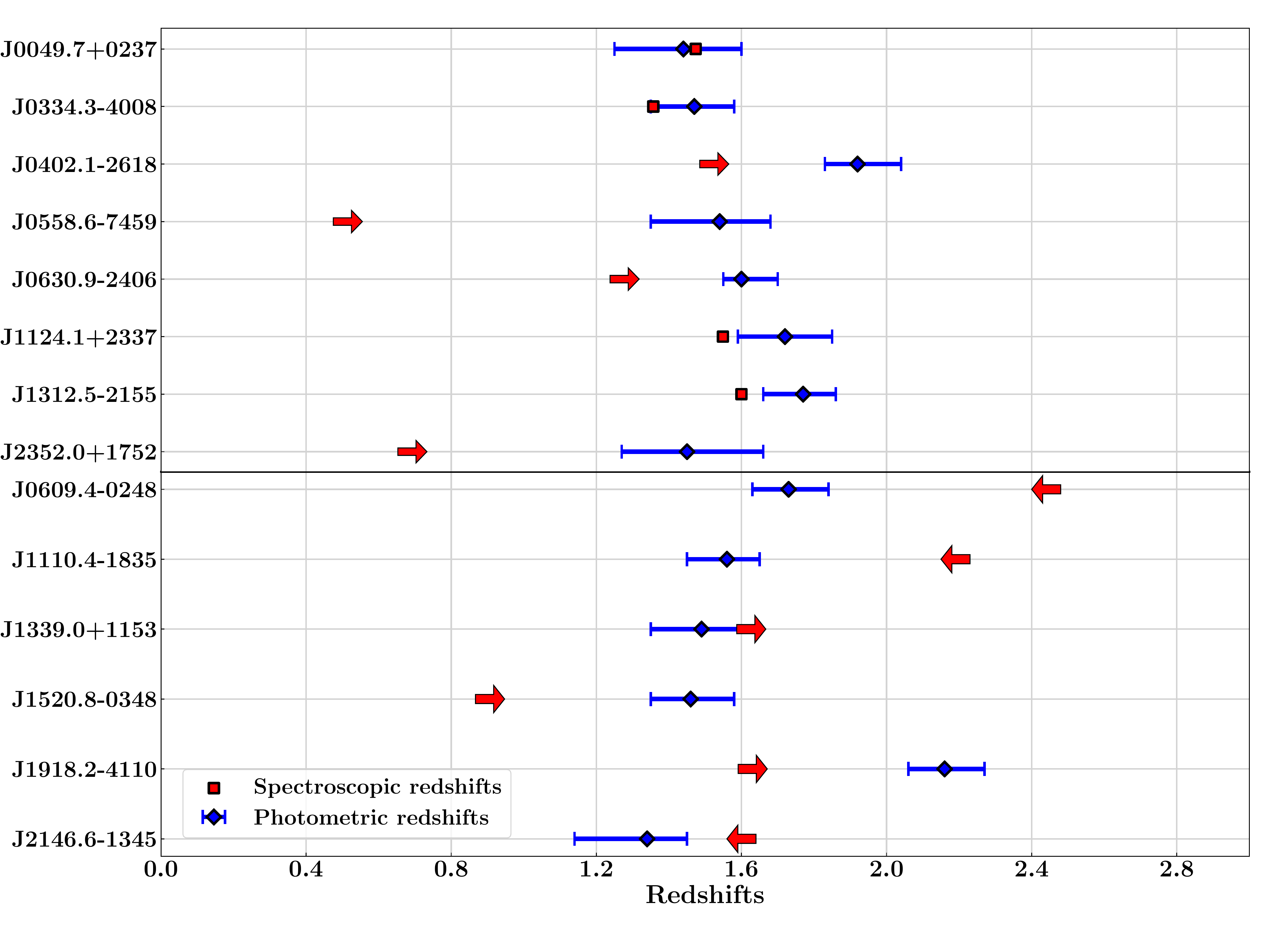}
\caption{\label{fig:specz} Photo-$z$ redshifts vs Spectroscopic redshifts for all high-$z$ sources found using this technique. The high-$z$ sources found by \cite{Rau2012} have been shown in the top panel, while the bottom panel shows sources from \cite{Kaur2017,Kaur2018} and the current work. The blue diamonds show the photometric redshift values, while the red squares are the confirmed spectroscopic redshifts from. All spectroscopic measurements have been obtained from \cite{shaw2013}.}
\end{figure}

The results from the SED fits performed for all 45 sources are listed in Table~\ref{tab:result}. We report here the photometric redshifts, the $P_z$ values, $\chi^2$ values and the best-fit models for power law and galaxy templates. None of the sources in our sample required a stellar template. \\
Based on the criteria mentioned in the previous section ($z>1.3$, $P_z > 90 \%$ and E(B-V)$\leq0.30$), we found 3 sources: 3FGL J0427.3$-$3900, 3FGL J0609.4$-$0248 and 3FGL J1110.4$-$1835, at redshifts greater than 1.3, which is consistent with the success rate of $\sim$10\% for this technique observed in \cite{Rau2012} and \cite{Kaur2017, Kaur2018}. The photometric redshifts for these 3 sources were found to be $z=1.42^{+0.12}_{-0.10}$, $z=1.73^{+0.11}_{-0.10}$ and $z=1.56^{+0.09}_{-0.11}$, respectively, using the power-law templates. The galaxy library which assumes hybrid QSO templates from \cite{Salvato2009, Salvato2011} yielded $z$ values of $1.25^{+0.10}_{-0.10}$, $1.34^{+0.09}_{-0.06}$ and $1.42^{+0.05}_{-0.05}$, respectively. But since these sources have been identified as BCU II (3FGL J0427.3$-$3900) and BL Lacs (3FGL J0609.4$-$0248 and 3FGL J1110.4$-$1835) in the 3LAC, the redshifts determined by the galaxy templates with dominant broad emission lines are unlikely. Hence, we base our analysis on the $z$ values derived from the power law templates. \\ 
In Figure~\ref{fig:sed}, SARA and $\emph{Swift}$-UVOT SEDs for these 3 sources are shown. 
For 17 sources from our sample, 90\% photometric redshift upper limits have been provided, while for 9 sources with E(B-V)$>0.3$, only the photometry results have been reported. For the remaining 16 sources, no $z_{\rm phot,best}$ is given as no satisfactory fit ($\chi^{2}>30$) was obtained. \\
\cite{shaw2013} reported the optical spectra of the 2 high-$z$ BL Lacs found here including the upper limits in their spectroscopic redshift measurements. Both sources exhibit featureless optical spectra typically observed for BL Lacs. The source 3FGL J0609.4$-$0248 is reported to have a spectroscopic upper limit of $z=2.48$ while 3FGL J1110.4$-$1835 has an upper limit of $z=2.23$, both consistent with our photometric findings.\\ 
In Figure~\ref{fig:specz}, the redshifts for BL Lacs obtained using the photometric technique have been compared to the available spectroscopic data. In most cases, spectroscopy is only able to place a lower or an upper limit for the redshift. As shown, redshifts obtained by the photometric technique are generally in good agreement with spectroscopic measurements. 
For 3FGL J1312.5-2155, the authors \citep{shaw2013} used a plausible Mg II feature to perform a single line identification in a small allowed redshift range (z $\sim$1.6). These data highlight the efficacy of the photometric method for determining the redshifts of high-$z$ BL Lacs.


\section{Discussion and Conclusions}\label{sec:dis}
We report 3 sources with $z>1.3$, bringing the total number of high-$z$ BL Lac sources to 29. Of these, 16 ($\sim$55\%) were discovered using the photometric dropout technique reported here. Thus, the photometric technique has successfully identified more than $50\%$ of the overall 29 high-$z$ BL Lacs known to date.
\begin{figure}[ht!]
\hspace{-0.6cm}
	\includegraphics[width=1.05\columnwidth,trim=10 0 0 0,clip=true]{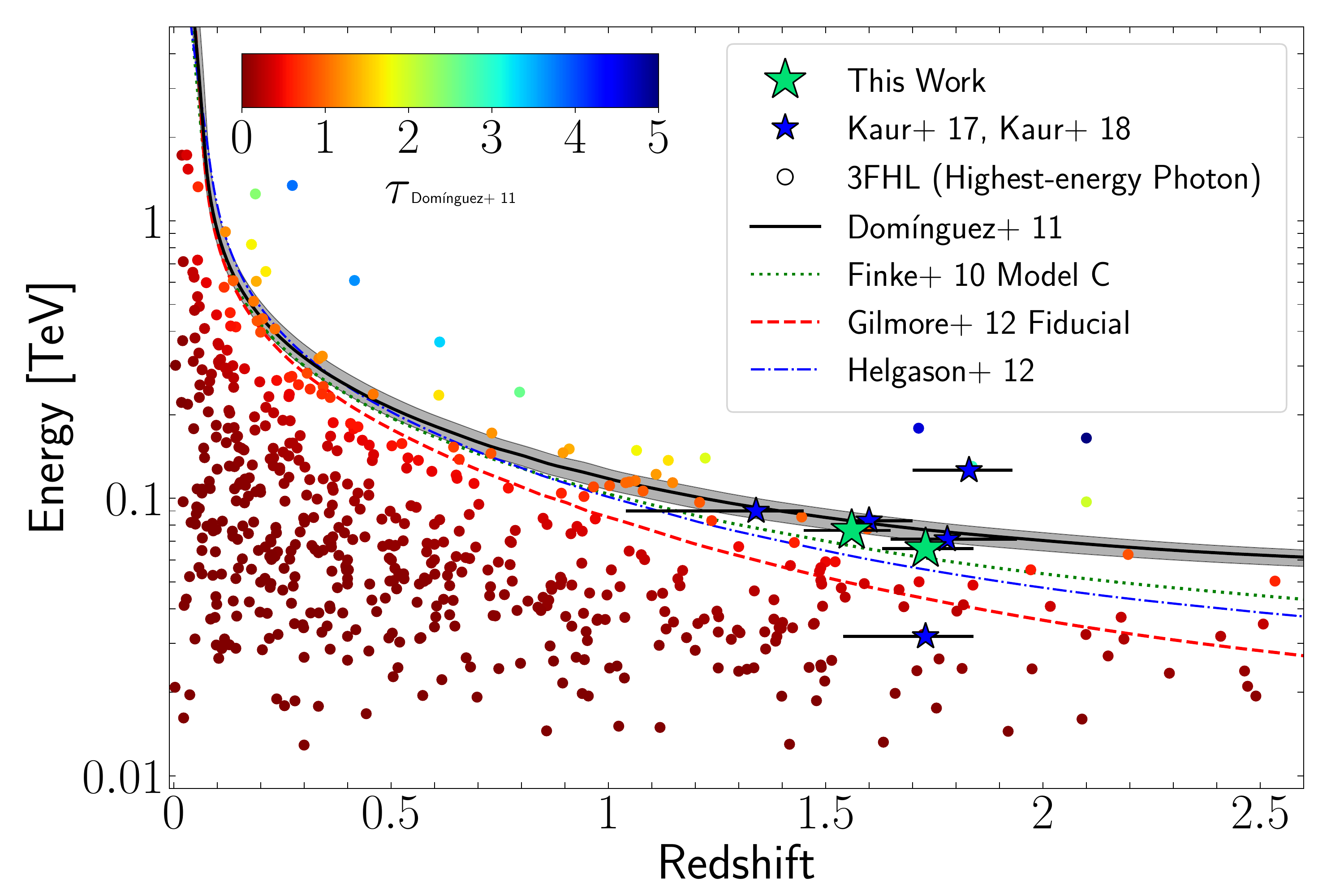}
	 \caption{\label{fig:cghp}The cosmic $\gamma$-ray horizon plot: The highest energy of photons from sources detected in the 3FHL catalog (E $>$ 10\,GeV; \citealp{3fhl}) vs. their redshift. The colors of points imply their corresponding optical depth ($\tau$) values (see colorbar). Various estimates of the cosmic $\gamma$-ray horizon, obtained from the EBL models by \cite{Finke2010} (dotted green line), \cite{Dominguez2011} (solid black line) with uncertainties as shaded band, \cite{Gilmore2012} (dashed red line) and \cite{Helgason2012} (dot-dashed blue line) are plotted for comparison. The highest energy photons from both the high-$z$ sources lie at the cosmic $\gamma$-ray horizon (green filled stars), whereas the other sources lie mostly below this limit. The blue star symbols represent the high-$z$ BL Lacs found in \citet{Kaur2017} and \citet{Kaur2018}.}
\end{figure}

\subsection{Cosmic $\gamma$-ray Horizon}
\begin{figure}[htp!]
\centering
\includegraphics[width=1.01\columnwidth]{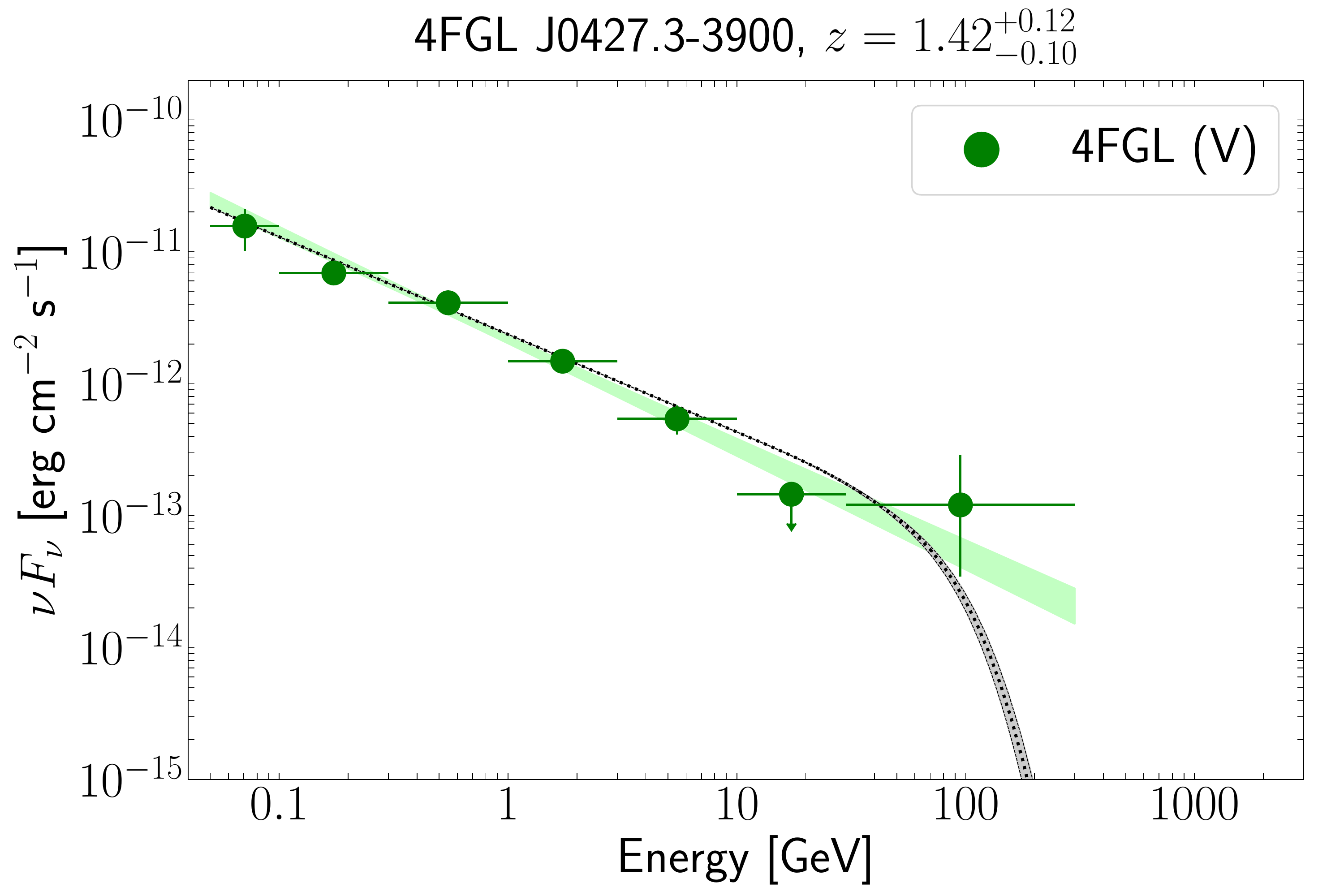}
\includegraphics[width=1.01\columnwidth]{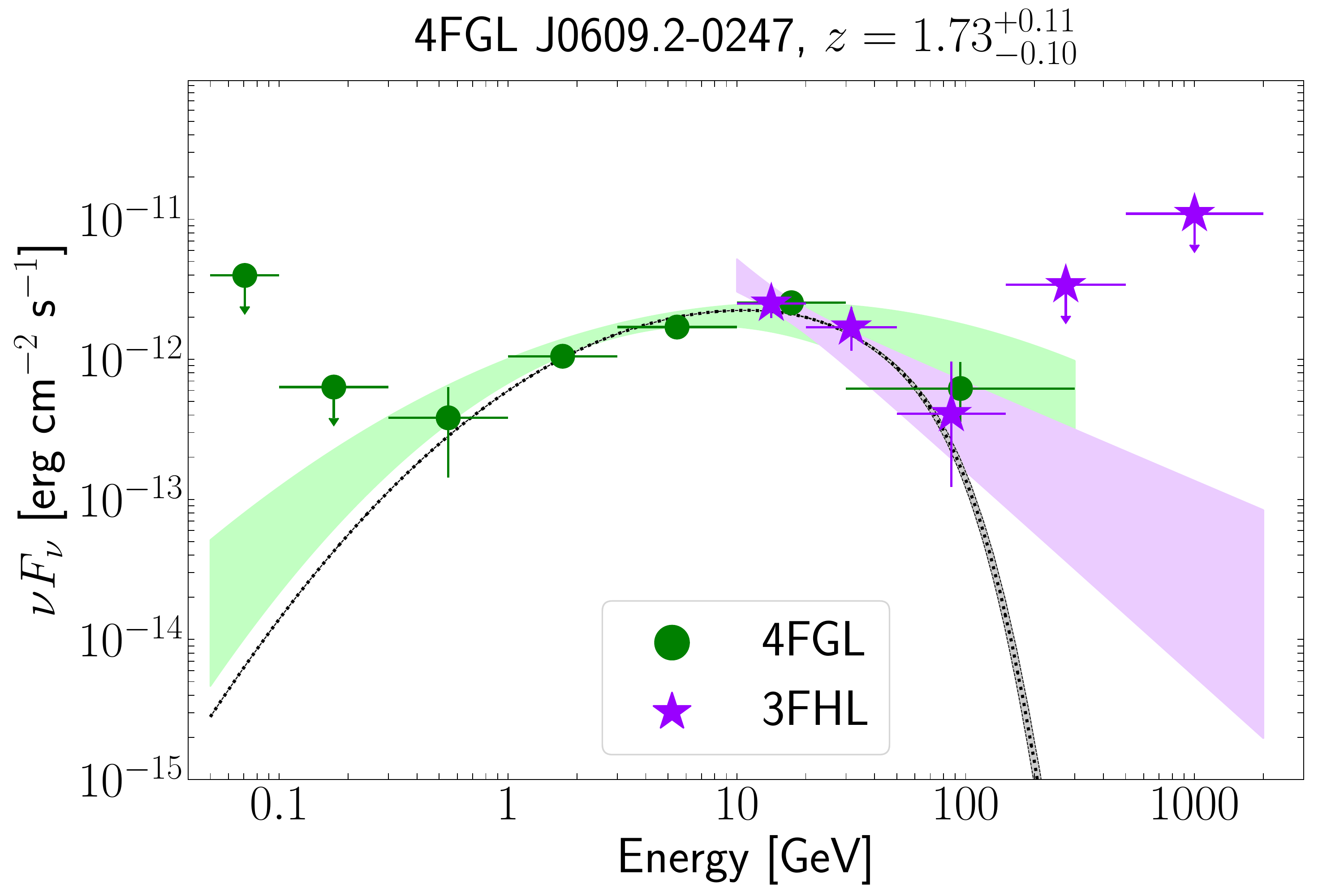}
\includegraphics[width=1.01\columnwidth]{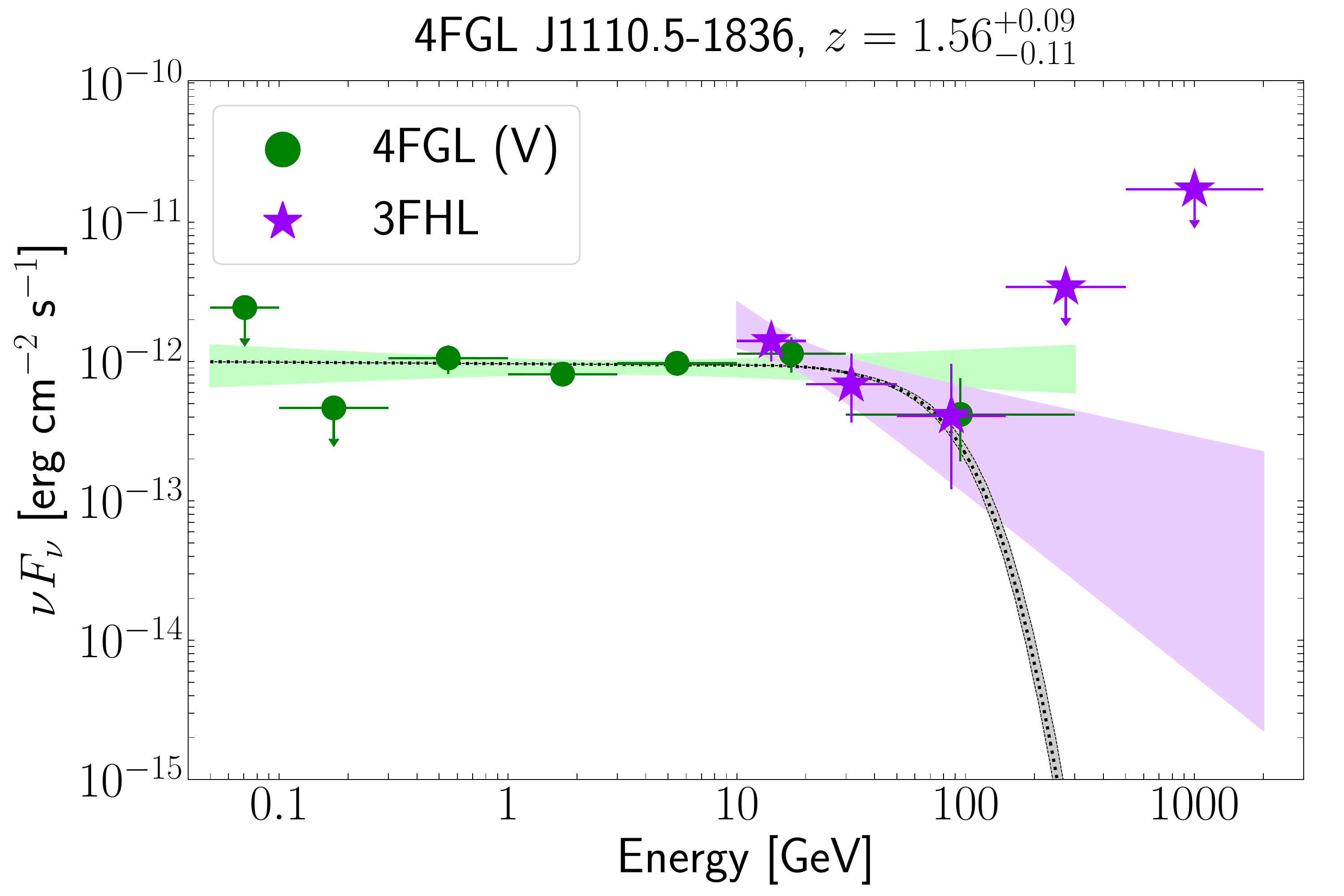}
\caption{\label{fig:seds}Spectral energy distributions of J0427.3$-$3900, J1110.4$-$1835 and J0609.4$-$0248 obtained by combining data reported in 3FHL and 4FGL \citep{3fhl,4fgl}. For the sources J0427.3$-$3900 and J1110.4$-$1835, the black line indicates a fit to the data with a power law absorbed by the EBL using the model by \citealp{Dominguez2011}. For the source J0609.4$-$0248, a log parabola fit is used to provide a better fit along with the EBL contribution. The grey region denotes uncertainty in the source flux that arises due to uncertainties in the redshift measurements. The green and purple regions denote the 1$\sigma$ uncertainty in 4FGL and 3FHL fits.}
\end{figure}

The cosmic $\gamma$-ray horizon (CGRH) provides us an estimate of the redshift at which the Universe becomes opaque to very high energy (VHE) $\gamma$-ray photons due to $\gamma$-$\gamma$ pair production with the EBL \citep{dominguez2013}. Quantifying constraints on the EBL requires sources with photons near or beyond the CGRH. In Figure~\ref{fig:cghp}, the CGRH has been plotted as a function of redshift for all the sources from the 3FHL catalog (colored circles; \citealp{3fhl}). The two high-$z$ sources we have found here (green stars) have also been included\footnote{The highest-energy photon for the source 3FGL J0427.3$-$3900 has not been reported in 2FHL/3FHL catalogs.} in the plot. The highest-energy photons of the two new sources are at 65.8 GeV (3FGL J0609.4$-$0248; $z$=1.73) and 76.6 GeV (3FGL J1110.4$-$1835; $z$=1.56), placing them right at the horizon. Thus, with this technique, we are able to locate sources that can be used to constrain the CGRH since they lie at a region where the optical depth, $\tau$, is between $1.2-1.8$, making them extremely valuable in these studies where redshift data are scarce. Providing a redshift for these sources will enable better and more accurate measurement of the EBL. \\
In Figure~\ref{fig:seds}, the $\gamma$-ray SEDs of the 3 high-$z$ sources are shown, constructed using data from the 3FHL and 4FGL catalogs \citep{3fhl,4fgl}. Two of these sources were fitted with power law models and one with a log parabola model to improve fit quality. EBL absorption was applied to all the sources by multiplying the spectral model by $e^{-\tau(E,z)}$, where $\tau$ is the optical depth due to the EBL as provided by \cite{Dominguez2011} model. Attenuation due to the EBL is apparent for all sources in the plot. Also apparent is the uncertainty due to the uncertain redshifts on the location of the EBL cutoff.

\begin{figure}[ht!]
\hspace{-0.6cm}
	\includegraphics[width=\columnwidth]{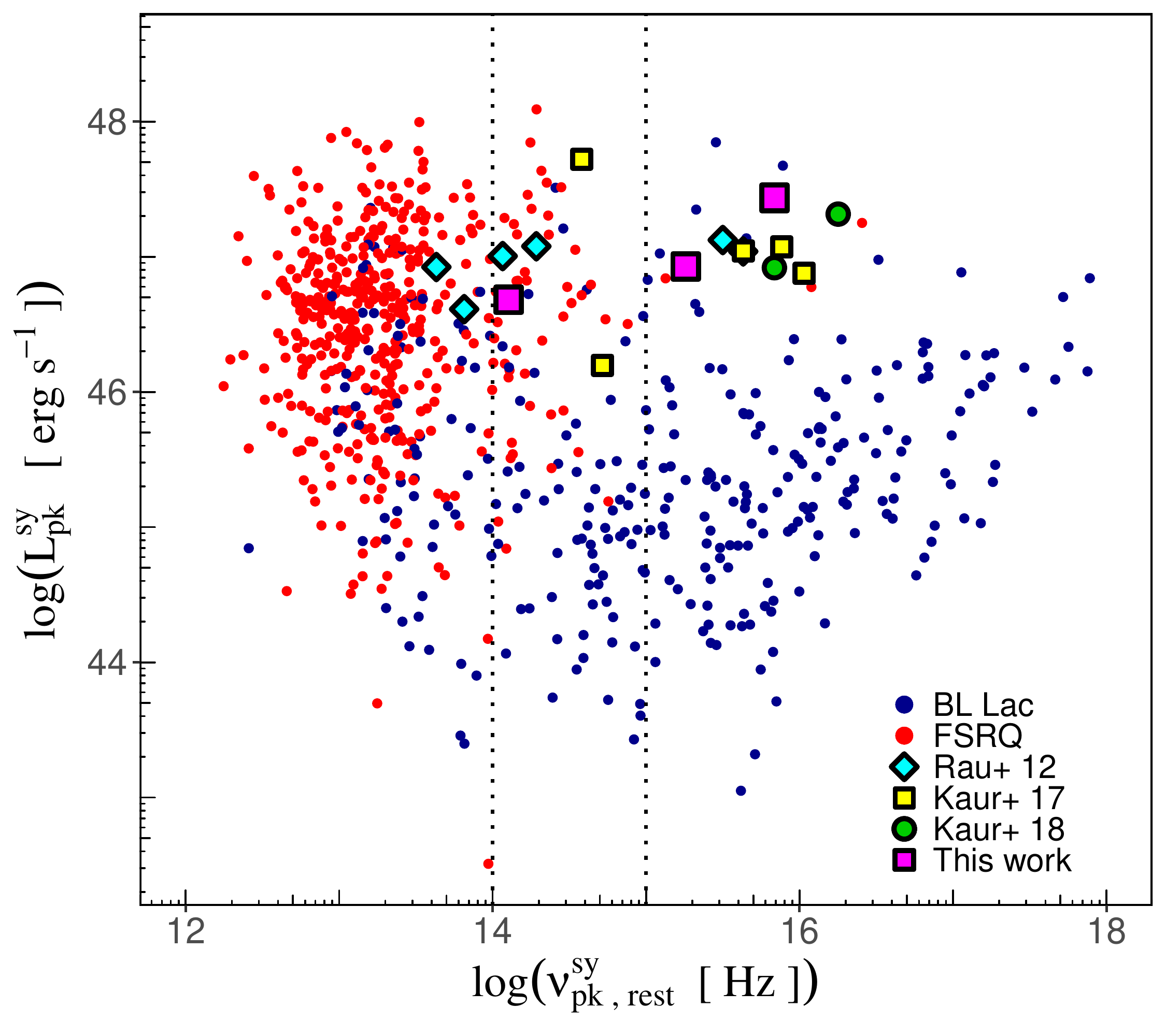}
			\caption{\label{fig:vlv}The peak synchrotron frequency, $\nu_{sy}^{pk}$ (in the rest frame) vs the peak synchrotron luminosity, $L_{sy}^{pk}$. The magenta stars represent the three new sources found in this work. They occupy a similar locus as other high-$z$ BL Lacs found using the photometric method. These sources display high luminosity and high synchrotron peak behavior. Six new BL Lacs from \citet{Rau2012} are represented by cyan diamonds, 5 BL Lacs from \citet{Kaur2017} in yellow squares, 2 BL Lacs found in \citet{Kaur2018} in green circles and all the FSRQs and BL Lacs from the 3LAC catalog with known redshifts are in red and blue circles, respectively. The separation of the LSP, ISP, and HSP regions are plotted using dotted vertical lines.}
\end{figure}
\subsection{The Blazar Sequence and The Fermi Blazar Divide}
The ``blazar sequence", introduced by \citet{fossati1998} (and later improved upon by \citealp{ghis2017}), suggests the existence of an anti-correlation between the synchrotron peak frequency ($\nu_{sy}^{pk}$) and the bolometric luminosity in order to provide a unified model for all blazar classes (an alternate view by \citet{Giommi2002, Padovani2002} suggests this could be a result of a selection effect). It was reported that blazars became redder (shifting towards lower frequencies) with increasing luminosity and at the same time, the Compton Dominance (CD, the ratio between the luminosities of the inverse Compton and synchrotron peaks) also increases. \\
FSRQs have been typically observed to have high luminosities, low synchrotron peak frequencies ($\nu_{sy}^{pk}\leq10^{13}$Hz) and CD $>1$, whereas BL Lacs are generally less luminous with high synchrotron peak frequencies and CD $\lesssim$ 1. These BL Lacs were also found to be less luminous in $\gamma$-rays ($L_{\gamma} \leq 10^{47}$ erg s$^{-1}$) and to possess harder spectra ($\Gamma_{\gamma} \leq 2.2$) than their FSRQ counterparts on the $\Gamma_{\gamma}-L_{\gamma}$ plane by \cite{ghis2009}. This ``Fermi Blazar Divide" was inferred as being due to FSRQs having higher accretion rates than BL Lacs. Therefore, according to this sequence, high-$z$ BL Lacs that are highly luminous (as FSRQs) and have $\nu_{sy}^{pk}$ $>10^{15}$ Hz should not exist.  \\
\begin{figure}[ht!]
\hspace{-0.6cm}
	\includegraphics[width=\columnwidth]{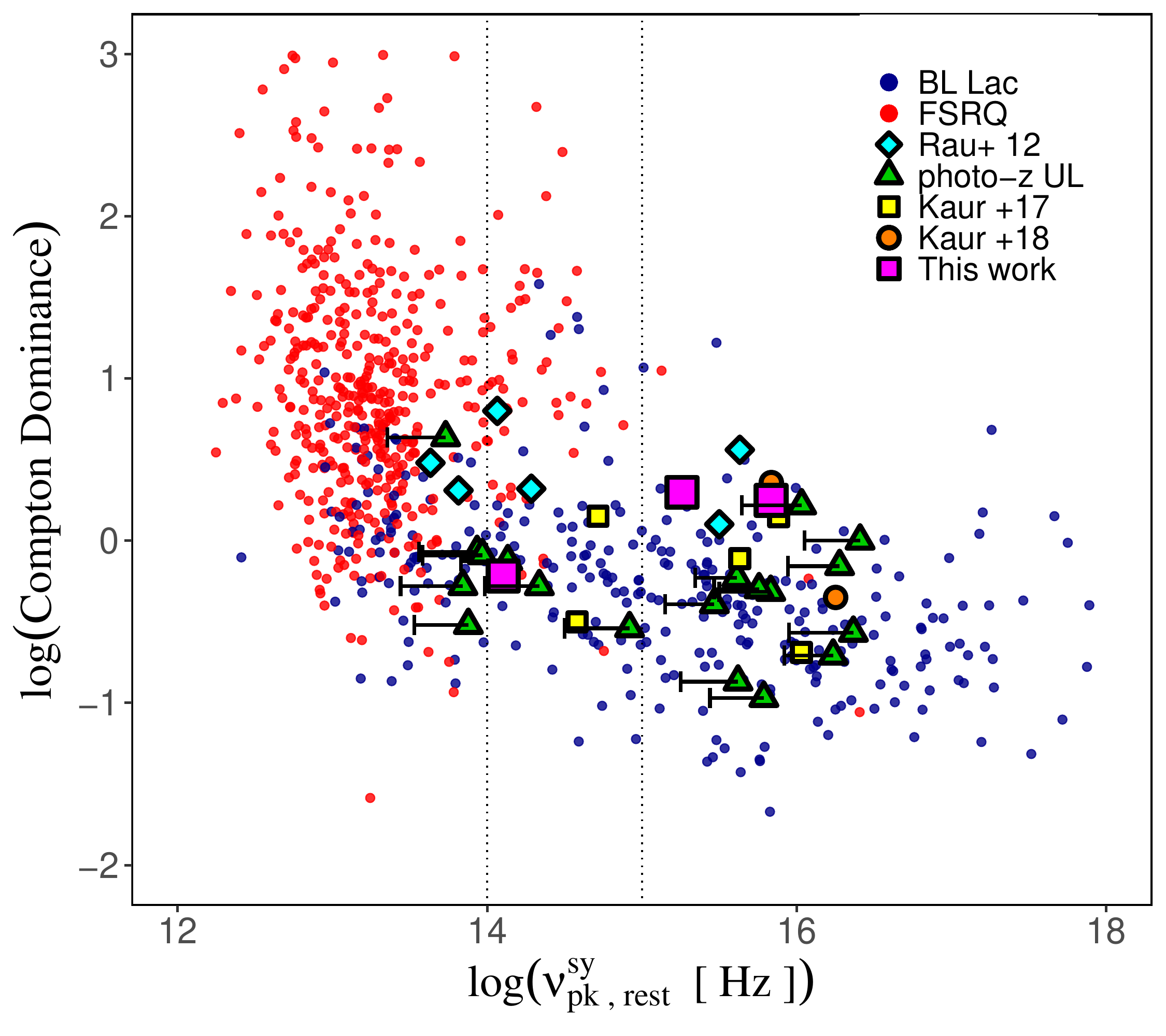}
			\caption{\label{fig:comp}Compton Dominance vs. $\nu_{sy}^{\rm pk}$. Upper limits are represented by green filled triangles and the 2 BL Lacs found in \citet{Kaur2018} are plotted in orange circles. The color scheme for the rest of the data symbols follows from Figure~\ref{fig:vlv}. }
\end{figure}

\begin{figure}[ht!]
\hspace{-0.6cm}
	\includegraphics[width=\columnwidth]{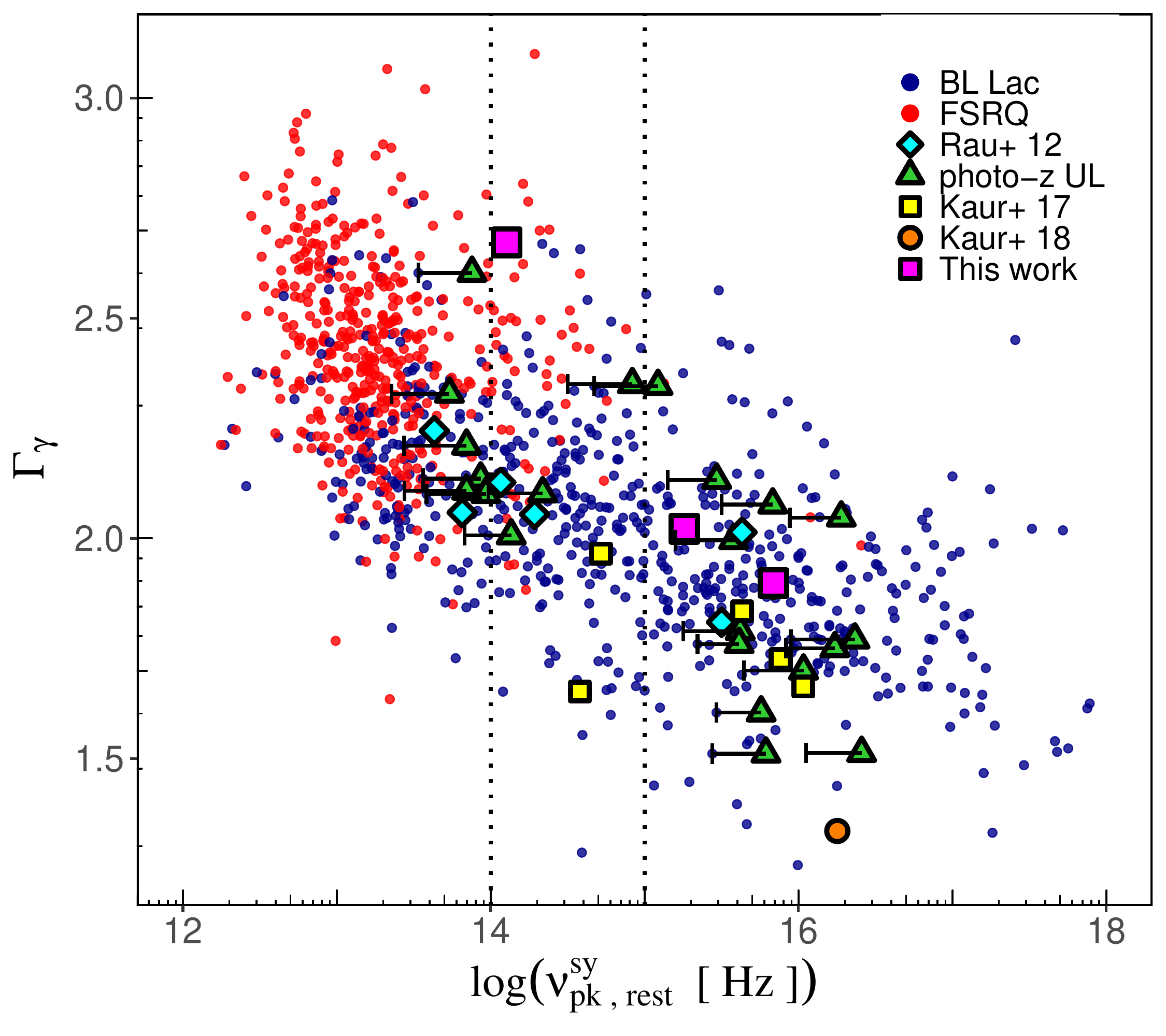}
			\caption{\label{fig:vgv}$\gamma$-ray photon index vs. rest-frame peak synchrotron frequency for all sources. The color scheme follows that used in Figure~\ref{fig:comp}.}
\end{figure}
\begin{figure}[ht!]
	\includegraphics[width=\columnwidth]{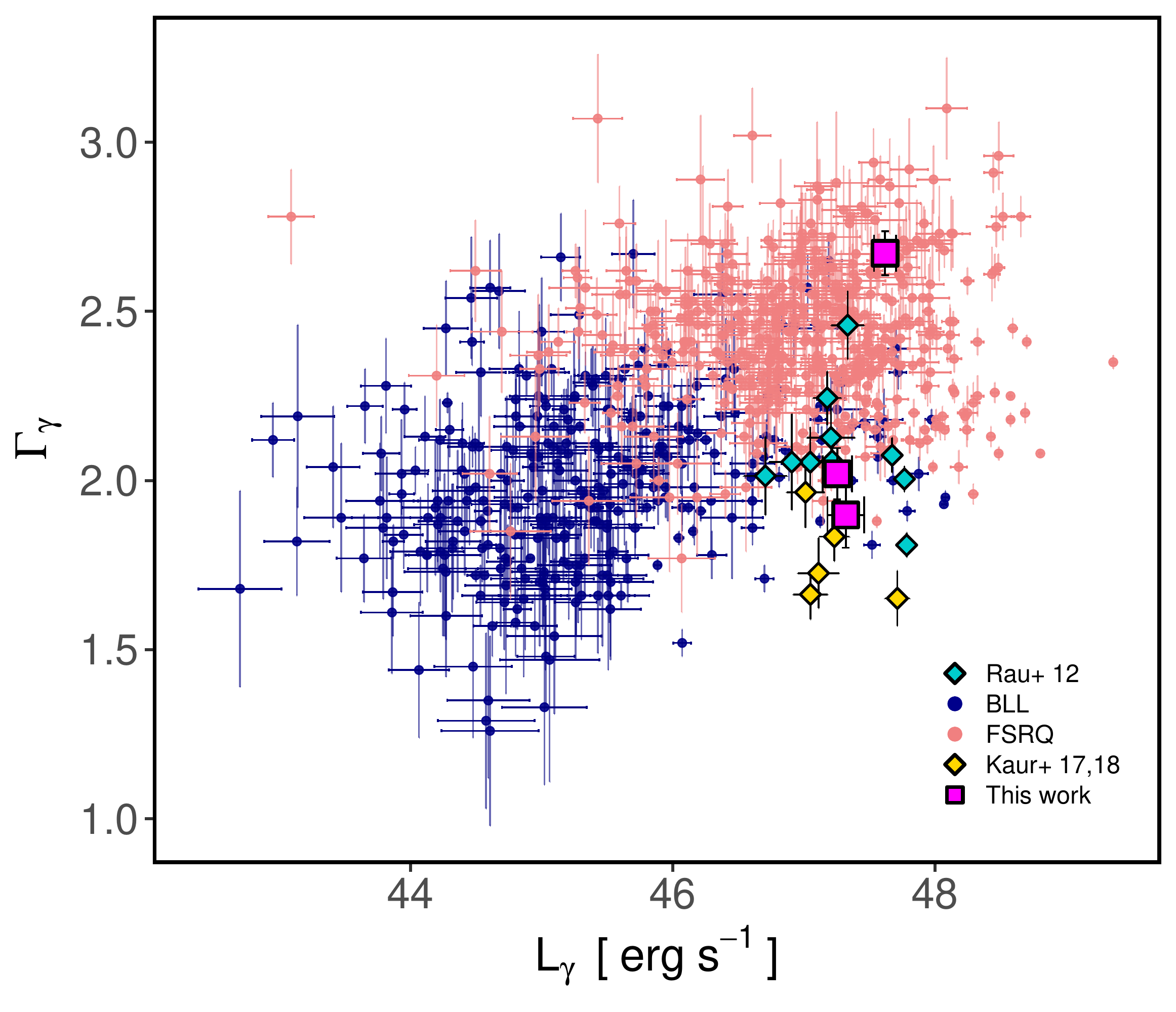}
			\caption{\label{fig:glg}Correlation between $\gamma$-ray luminosity and the spectral index.}
\end{figure}
We plot these correlations using blazars from the 3LAC catalog and all the high-$z$ sources found so far using the photometric technique. The values for $\nu_{sy}^{pk}$, $L_{sy}^{pk}$ and the $\gamma$-ray photon index were extracted from the 3LAC catalog and Compton Dominance values were calculated with the help of the online SED tool\footnote{\url{https://tools.ssdc.asi.it/}}. These plots are represented in Figures~\ref{fig:vlv} - \ref{fig:glg}.\\
Figure~\ref{fig:vlv} shows that majority of our sources lie in the region where both $\nu_{sy}^{pk}$ and $L_{sy}^{pk}$ are relatively high, in direct contradiction with the blazar sequence. The sources discussed here confirm the previously established anti-correlation of CD with $\nu_{sy}^{pk}$ (where CD $\lesssim$ 1) and have hard $\gamma$-ray spectra, typical of BL Lacs (Figures~\ref{fig:comp}, \ref{fig:vgv}). Two of the three high-$z$ BL Lacs detected in this work have hard spectra ($\Gamma_{\gamma} \leq 2$) and $\gamma$-ray luminosities, $L_{\gamma} \geq 10^{47}$ erg s$^{-1}$ (see Figure~\ref{fig:glg}). Even though the $L_{\gamma}$ of these sources are similar to those seen for FSRQs, the spectral indices are in agreement with those of BL Lacs. \\
Division based on the mass accretion rates for blazars has been extensively argued by various authors \citep{ghis2012, padovani2012} who propose these candidates to be ``masquerading BL Lacs", i.e. FSRQs whose broad emission lines have been swamped by the relativistic non-thermal continuum and are therefore hidden by the bright synchrotron emission peaking in the UV region. The high synchrotron peak frequencies ($\nu_{sy}^{pk}>10^{14}-10^{15}$ Hz) in these sources imply that the emission region lies further out and suffers less cooling via the external Compton process. 
However, this scenario is still being debated since identifying such objects is extremely difficult \citep{raj2020}. Recently, TXS 0506+056, the first cosmic non-stellar neutrino source reported by the \citet{icecube2018} has been found to be a masquerading BL Lac instead of a BL Lac (see e.g. \citealp{padovani2019}).\\
The average mass for {\it Fermi}-LAT detected FSRQs was found to be $5\times10^8$\Msun by \cite{sbarr2012} and an average mass of $8\times10^8$\Msun was obtained by \cite{paliya_cgrabs} for radio-loud CGrabs objects \citep{2008ApJS..175...97H}. Considering these as the black hole mass limits for our high-$z$ sources, we can obtain the BLR luminosity, $L_{BLR}$, using the relation given by \cite{sbarr2012}:\,
\begin{equation}
L_{BLR} \sim 4L_{\gamma}^{0.93},
\end{equation}
yielding $L_{BLR} \sim$ $2\times10^{44}$ erg s$^{-1}$. Assuming that 10\% of the disk luminosity ($L_{d}$) is reprocessed by the BLR, we obtain $L_{d} \sim$ $2\times10^{45}$ erg s$^{-1}$. Using these values, the ratio between the disk luminosity and the Eddington luminosity (the maximum possible luminosity of a body that can be achieved when there is a balance between radiation force and gravitational force), $L_{d}/L_{Edd}$ can be obtained. Found to be in the range $L_{d}/L_{Edd}\sim 0.02-0.03$, these values are typically observed in case of powerful FSRQs with highly efficient accretion disks. Since the synchrotron peak frequencies of these sources are also observed to be bluer ($\nu_{sy}^{pk}>10^{14}-10^{15}$ Hz), these objects could belong to the elusive class of masquerading BL Lacs.\\
This and the previous studies employing the photometric technique significantly enlarged the blazar sample in the high $\nu_{sy}^{pk}$, high $L_{sy}^{pk}$ part of the phase diagram (Figure~\ref{fig:vlv}), which provides a powerful diagnostic for this class of blazars. Thus, a small but non-negligible population of high $\nu_{sy}^{pk}$, high $L_{sy}^{pk}$ sources is starting to emerge. However, the sample size remains small and further studies need to be undertaken in order to draw concrete conclusions.  

\balance
\acknowledgements
MR and MA acknowledge funding under NASA contract 80NSSC18K1717. They also acknowledge the {\it Swift} team for scheduling all the {\it Swift}-UVOT observations. Reported work is in part based on observations obtained with the SARA Observatory telescopes at Chile (SARA-CT) and La Palma (SARA-ORM), which are owned and operated by the Southeastern Association for Research in Astronomy (\href{www.saraobservatory.org}{saraobservatory.org}). More information about SARA can be found in DOI:   \href{https://iopscience.iop.org/article/10.1088/1538-3873/129/971/015002}{10.1088/1538-3873/129/971/015002}.
A.D. is thankful for the support of the Ram{\'o}n y Cajal program from the Spanish MINECO.\\


\clearpage
\startlongtable
\begin{deluxetable*}{llcccrc}
	\tablecolumns{7}
	\tablecaption{\label{tab:obs} $\emph{Swift}$-UVOT and SARA Observations along with visual extinction values. $A_{V}=3.1\times$E(B-V)}
	\tablewidth{0pt}
	\tabletypesize{\small}
	\setlength{\tabcolsep}{0.07in} 
	\tablehead{
		\colhead{3FGL }   & \colhead{$\emph{$\emph{Counterpart}$}$} & \colhead{RA J2000} &	\colhead{Dec J2000} &	\colhead{$\emph{Swift}$ Date$^{a}$} & 	\colhead{SARA Date$^{a}$} & 	\colhead{$A_{V}$} \\
		\colhead{ (Name)}   & \colhead{(Name)} & \colhead{(hh:mm:ss)} &	\colhead{($^\circ: ^\prime: ''$)} &	\colhead{(UT)} & 	\colhead{(UT)} & 	\colhead{(mag)}
    	}
\startdata
J0009.1$+$0630 & GB6 J0009$+$0625 & 00:09:17 & $+$06:25:54 & 2018$-$12$-$22 & 2018$-$12$-$08 & 0.17\\
J0013.2$-$3954 & PKS 0010$-$401 & 00:13:00 & $-$39:54:26 & 2019$-$01$-$05 & 2018$-$12$-$15 & 0.03\\
J0019.4$+$2021 & PKS0017$+$200 & 00:19:38 & $+$20:21:45 & 2018$-$09$-$20 & 2018$-$09$-$30 & 0.16\\
J0026.7$-$4603 & 1RXS J002636.3$-$460101 &  00:26:35 & $-$46:01:10 & 2018$-$12$-$18 & 2018$-$11$-$29 & 0.03\\
J0041.9$+$3639 & RX J0042.0$+$3641 & 00:42:08 & $+$36:41:12 & 2019$-$03$-$04 & 2018$-$11$-$25 & 0.12\\
J0116.3$-$6153 & SUMSS J011619$-$615343 & 01:16:19 & $-$61:53:43 & 2019$-$01$-$24 & 2018$-$11$-$29 & 0.05\\
J0125.4$-$2548 & PKS 0122$-$260 & 01:25:19 & $-$25:49:04 & 2019$-$01$-$17 & 2018$-$12$-$15 & 0.04\\
J0133.0$-$4413 & SUMSS J013306$-$441422 & 01:33:06 & $-$44:14:21 & 2018$-$12$-$29 & 2018$-$12$-$07 & 0.05\\
J0143.7$-$5845 & SUMSS J014347$-$584550 & 01:43:47 & $-$58:45:51 & 2018$-$12$-$07 & 2018$-$11$-$25 & 0.05\\
J0152.8$+$7517 & 1RXS J015308.4$+$751756 & 01:53:07 & $+$75:17:44 & 2019$-$03$-$07 & 2018$-$11$-$25 & 1.27\\
J0204.0$+$7234 & S5 0159$+$723\footnote{The presence of a 4th magnitude star in the field of this AGN prevented UVOT observations.} & 02:03:33 & $+$72:32:53 & $-$ & 2018$-$11$-$25 & 1.79\\
J0253.0$-$0125 & FBQS J0253$-$0124 & 02:53:15 & $-$01:24:05 & 2019$-$01$-$13 & 2018$-$12$-$15 & 0.16\\
J0304.3$-$2836 & RBS 0385 & 03:04:16 & $-$28:32:18 & 2018$-$11$-$21 & 2018$-$11$-$17 & 0.04\\ 
J0316.2$-$6436 & SUMSS J031614$-$643732 & 03:16:14 & $-$64:37:31 & 2019$-$02$-$02 & 2018$-$12$-$29 & 0.08\\
J0335.3$-$4459 & 1RXS J033514.5$-$445929 & 03:35:14 & $-$44:59:39 & 2019$-$02$-$15 & 2018$-$12$-$15 & 0.03\\
J0359.3$-$2612 & PKS 0357$-$264 & 03:59:33 & $-$26:15:31 & 2018$-$12$-$28 & 2018$-$11$-$29 & 0.08\\
J0409.8$-$0358 & NVSS J040946$-$040003 & 04:09:46 & $-$04:00:00 & 2018$-$12$-$12 & 2018$-$11$-$25 & 0.20\\
J0427.3$-$3900 & PMN J0427$-$3900 & 04:27:21 & $-$39:01:00 & 2017$-$11$-$23 & 2018$-$11$-$25 & 0.08\\
J0505.9$+$6114 & NVSS J050558$+$611336 & 05:05:58 & $+$61:13:36 & 2019$-$03$-$21 & 2018$-$12$-$01 & 1.65\\ 
J0529.2$-$5917 & 1RXS J052846.9$-$592000 & 05:28:47 & $-$59:20:00 & 2018$-$03$-$28 & 2018$-$04$-$01 & 0.06\\
J0556.0$-$4353 & SUMSS J055618$-$435146 & 05:56:18 & $-$43:51:46 & 2018$-$04$-$04 & 2018$-$11$-$29 & 0.16\\
J0609.4$-$0248 & NVSS J060915$-$024754 & 06:09:15 & $-$02:47:54 & 2018$-$04$-$11 & 2018$-$11$-$25 & 0.80\\
J0627.9$-$1517 & NVSS J062753$-$152003 & 06:27:53 & $-$15:20:03 & 2017$-$12$-$20 & 2017$-$12$-$22 & 0.53\\
J0807.1$-$0541 & PKS 0804$-$05 & 08:07:09 & $-$05:41:13 & 2018$-$05$-$09 & 2018$-$12$-$01 & 0.10\\
J0958.3$-$0318 & 1RXS J095806.4$-$031729 & 09:58:06 & $-$03:17:38 & 2019$-$04$-$07 & 2019$-$03$-$27 & 0.09\\
J1104.3$+$0730 & MG1 J110424$+$0730 & 11:04:24 & $+$07:30:53 & 2018$-$06$-$06 & 2018$-$06$-$17 & 0.11\\
J1110.4$-$1835 & CRATES J111027.78$-$183552.6 & 11:10:27 & $-$18:35:52 & 2018$-$04$-$19 & 2018$-$12$-$29 & 0.12\\
J1224.6$-$8312 & PKS 1221$-$82 & 12:24:54 & $-$83:13:10 & 2018$-$01$-$01 & 2018$-$04$-$12 & 0.68\\
J1304.3$-$5535 & PMN J1303$-$5540 & 13:03:49 & $-$55:40:31 & 2018$-$02$-$15 & 2018$-$03$-$20 & 1.13\\
J1326.6$-$5256 & PMN J1326$-$5256 & 13:26:49 & $-$52:56:23 & 2018$-$06$-$22 & 2018$-$06$-$17 & 1.07\\
J1353.5$-$6640 & 1RXS J135341.1$-$664002 & 13:53:35 & $-$66:40:06 & 2018$-$06$-$22 & 2018$-$08$-$28 & 1.48\\
J1427.6$-$3305 & PKS B1424$-$328 & 14:27:41 & $-$33:05:31 & 2018$-$09$-$11 & 2019$-$03$-$27 & 0.17\\
J1440.4$-$3845 & 1RXS J144037.4$-$384658 & 14:40:37 & $-$38:46:53 & 2018$-$05$-$18 & 2019$-$03$-$27 & 0.25\\
J1508.7$-$4956 & ICRF J150838.9$-$495302 & 15:08:39 & $-$49:53:02 & 2018$-$02$-$01 & 2018$-$04$-$12 & 1.00\\
J1509.9$-$2951 & TXS 1507$-$296 & 15:10:09 & $-$29:51:34 & 2018$-$03$-$11 & 2018$-$03$-$20 & 0.56\\
J1525.2$-$5905 & PMN J1524$-$5903 & 15:24:51 & $-$59:03:39 & 2018$-$03$-$11 & 2018$-$04$-$01 & 5.77\\
J1645.2$-$5747 & AT20G J164513$-$575122 & 16:45:13 & $-$57:51:22 & 2018$-$03$-$15 & 2018$-$03$-$20 & 0.76\\
J1745.4$-$0754 & TXS 1742$-$078 & 17:42:44 & $-$07:51:54 & 2018$-$07$-$11 & 2018$-$08$-$28 & 2.42\\
J2200.2$+$2139 & TXS 2157$+$213 & 22:00:14 & $+$21:37:57 & 2019$-$04$-$08 & 2018$-$11$-$25 & 0.27\\
J2227.8$+$0040 & PMN J2227$+$0037 & 22:27:57 & $+$00:37:03 & 2018$-$07$-$17 & 2018$-$11$-$29 & 0.16\\
J2247.8$+$4413 & NVSS J224753$+$441317 & 22:47:53 & $+$44:13:15 & 2019$-$02$-$23 & 2018$-$11$-$25 & 0.65\\ 
J2307.4$-$1208 & 1RXS J230722.5$-$120520 & 23:07:22 & $-$12:05:18 & 2018$-$11$-$29 & 2018$-$12$-$01 & 0.09\\
J2319.2$-$4207 & PKS 2316$-$423 & 23:19:06 & $-$42:06:48 & 2018$-$11$-$04 & 2018$-$11$-$29 & 0.05\\
J2324.7$+$0801 & PMN J2324$+$0801 & 23:24:45 & $+$08:02:06 & 2018$-$11$-$11 & 2018$-$11$-$25 & 0.24\\
J2357.4$-$1716 & RBS 2066 & 23:57:30 & $-$17:18:03 & 2018$-$07$-$08 & 2018$-$11$-$15 & 0.06\\
\enddata
\tablenotetext{a}{The observation dates for for $Swift$ and SARA correspond to the beginning of the exposures.}
    \end{deluxetable*}
    

\begin{center}
\begin{longrotatetable}
\begin{deluxetable*}{llllllrrrrr}
	\tablecolumns{11}
	\tablecaption{\label{tab:mags} $\emph{Swift}$-UVOT and SARA photometry (AB magnitudes corrected for extinction)}
	\tablewidth{0pt}
	\tabletypesize{\footnotesize}
	\setlength{\tabcolsep}{0.03in} 
	\tablehead{
\colhead{3FGL Name}   & \colhead{$g^\prime$} & \colhead{$r^\prime$} &	\colhead{$i^\prime$} &	\colhead{$z^\prime$} & 	\colhead{$uvw2$} & 	\colhead{$uvm2$} & 	\colhead{$uvw1$} & 	\colhead{$u$} & \colhead{$b$} & \colhead{$v$}}
	\startdata
J0009.1$+$0630 & 20.47$\pm$0.13 & 19.69$\pm$0.05 & 19.81$\pm$0.09 & 19.91$\pm$0.26 & 22.11$\pm$0.24 & 21.66$\pm$0.24 & $>$21.4 & 21.05$\pm$0.28 & 20.37$\pm$0.31 & $>$19.4\\
J0013.2$-$3954 & 19.59$\pm$0.10 & 18.73$\pm$0.03 & 18.59$\pm$0.05 & 18.56$\pm$0.10 & 20.68$\pm$0.14 & 20.27$\pm$0.27 & 20.15$\pm$0.18 & 19.44$\pm$0.14 & 19.74$\pm$0.32 & 18.36$\pm$0.26\\
J0019.4$+$2021 & 19.93$\pm$0.03 & 19.07$\pm$0.02 & 18.65$\pm$0.02 & 18.88$\pm$0.04 & $>$22.1 & $>$21.7 & $>$21.3 & $>$20.8 & $>$20.1 & $>$19.2\\
J0026.7$-$4603 & 18.06$\pm$0.03 & 17.52$\pm$0.02 & 17.30$\pm$0.03 & 17.57$\pm$0.07 & 18.43$\pm$0.07 & 18.25$\pm$0.09 & 18.36$\pm$0.11 & 18.12$\pm$0.11 & 18.15$\pm$0.16 & 17.88$\pm$0.23\\
J0041.9$+$3639 & 18.87$\pm$0.02 & 18.45$\pm$0.01 & 18.31$\pm$0.01 & 18.44$\pm$0.02 & 19.67$\pm$0.11 & 19.53$\pm$0.12 & 19.18$\pm$0.15 & 19.40$\pm$0.30 & $>$18.9 & $>$18.2\\
J0116.3$-$6153 & 18.83$\pm$0.05 & 18.24$\pm$0.03 & 18.00$\pm$0.04 & 17.71$\pm$0.06 & 19.68$\pm$0.13 & 19.50$\pm$0.15 & 19.33$\pm$0.22 & 19.13$\pm$0.25 & 18.93$\pm$0.26 & $>$18.3\\
J0125.4$-$2548 & 21.99$\pm$0.57 & 21.24$\pm$0.31 & 19.58$\pm$0.22 & 19.76$\pm$0.28 & $>$24.4 & $>$23.9 & $>$23.3 & $>$22.8 & $>$22.1 & $>$21.4\\
J0133.0$-$4413 & 19.37$\pm$0.05 & 18.41$\pm$0.02 & 17.94$\pm$0.03 & 17.61$\pm$0.05 & 20.55$\pm$0.13 & 20.39$\pm$0.18 & 20.35$\pm$0.19 & 20.32$\pm$0.27 & 19.48$\pm$0.23 & $>$19.2\\
J0143.7$-$5845 & 17.28$\pm$0.02 & 16.94$\pm$0.01 & 16.87$\pm$0.01 & 16.63$\pm$0.03 & 17.26$\pm$0.07 & 17.25$\pm$0.08 & 17.14$\pm$0.08 & 17.02$\pm$0.08 & 17.34$\pm$0.15 & 16.66$\pm$0.14\\
J0152.8$+$7517 & 17.21$\pm$0.02 & 16.92$\pm$0.01 & 16.86$\pm$0.01 & 17.12$\pm$0.02 & 17.77$\pm$0.30 & 18.06$\pm$0.36 & 17.36$\pm$0.28 & $>$17.8 & 17.25$\pm$0.29 & 16.71$\pm$0.25\\
J0204.0$+$7234 & 20.46$\pm$0.20 & 18.9$\pm$0.07 & 18.23$\pm$0.05 & 21.57$\pm$2.18 & \nodata & \nodata & \nodata & \nodata & \nodata & \nodata\\
J0253.0$-$0125 & 19.89$\pm$0.13 & 18.87$\pm$0.04 & 17.25$\pm$0.07 & 19.09$\pm$0.25 & 20.65$\pm$0.17 & 20.26$\pm$0.16 & 20.11$\pm$0.19 & 19.72$\pm$0.19 & $>$19.8 & $>$18.9\\
J0304.3$-$2836 & 20.31$\pm$0.05 & 18.88$\pm$0.03 & 18.50$\pm$0.04 & 18.47$\pm$0.08 & 21.34$\pm$0.13 & 21.27$\pm$0.16 & 21.27$\pm$0.19 & 20.97$\pm$0.20 & 20.58$\pm$0.24 & $>$20.2\\
J0316.2$-$6436 & 18.31$\pm$0.31 & 17.80$\pm$0.02 & 17.60$\pm$0.02 & 17.46$\pm$0.05 & 18.47$\pm$0.06 & 18.52$\pm$0.08 & 18.32$\pm$0.07 & 18.34$\pm$0.08 & 18.29$\pm$0.10 & 18.04$\pm$0.16\\
J0335.3$-$4459 & 18.27$\pm$0.03 & 17.62$\pm$0.01 & 15.75$\pm$0.2 & 17.12$\pm$0.04 & 19.10$\pm$0.05 & 18.86$\pm$0.07 & 18.65$\pm$0.07 & 18.44$\pm$0.06 & 18.34$\pm$0.08 & 17.73$\pm$0.09\\
J0359.3$-$2612 & 19.89$\pm$0.20 & 19.81$\pm$0.16 & 19.36$\pm$0.18 & 19.31$\pm$0.37 & $>$21.9 & $>$21.4 & $>$21.1 & $>$20.7 & $>$20.0 & $>$19.1\\
J0409.8$-$0358 & 16.81$\pm$0.02 & 16.17$\pm$0.01 & 16.04$\pm$0.01 & 15.92$\pm$0.02 & 18.11$\pm$0.08 & 17.93$\pm$0.10 & 17.59$\pm$0.09 & 17.43$\pm$0.09 & 16.92$\pm$0.09 & 16.56$\pm$0.12\\
J0427.3$-$3900 & 18.76$\pm$0.07 & 18.46$\pm$0.04 & 18.37$\pm$0.05 & 18.40$\pm$0.11 & 19.86$\pm$0.10 & 19.28$\pm$0.10 & 19.98$\pm$0.10 & 18.88$\pm$0.11 & 18.80$\pm$0.19 & 18.66$\pm$0.28\\
J0505.9$+$6114 & 17.89$\pm$0.03 & 17.48$\pm$0.01 & 17.33$\pm$0.01 & 17.64$\pm$0.03 & $>$18.1 & $>$18.3 & $>$17.9 & $>$18.2 & $>$17.9 & $>$17.7\\
J0529.2$-$5917 & 19.26$\pm$0.08 & 18.77$\pm$0.04 & 18.54$\pm$0.04 & 17.93$\pm$0.05 & $>$21.4 & $>$20.5 & $>$19.9 & $>$19.6 & $>$19.3 & $>$18.8\\ 
J0556.0$-$4353 & 19.21$\pm$0.13 & 18.87$\pm$0.08 & 18.53$\pm$0.08 & 18.33$\pm$0.14 & 20.51$\pm$0.13 & 20.10$\pm$0.13 & 19.98$\pm$0.17 & 19.83$\pm$0.22 & 19.27$\pm$0.22 & 19.31$\pm$0.34\\
J0609.4$-$0248 & 16.62$\pm$0.04 & 16.24$\pm$0.02 & 16.27$\pm$0.02 & 16.21$\pm$0.03 & 17.78$\pm$0.17 & 18.19$\pm$0.25 & 17.40$\pm$0.16 & 16.86$\pm$0.11 & 16.68$\pm$0.13 & 13.32$\pm$0.16\\
J0627.9$-$1517 & 19.83$\pm$0.12 & 15.07$\pm$0.01 & 18.16$\pm$0.03 & 18.09$\pm$0.06 & 21.89$\pm$0.28 & 21.74$\pm$0.28 & $>$21.9 & $>$21.8 & $>$21.2 & $>$20.5\\
J0807.1$-$0541 & 17.51$\pm$0.01 & 17.14$\pm$0.01 & 16.93$\pm$0.01 & 16.90$\pm$0.01 & 18.97$\pm$0.07 & 18.99$\pm$0.10 & 18.58$\pm$0.08 & 18.11$\pm$0.09 & 17.57$\pm$0.09 & 17.77$\pm$0.17\\
J0958.3$-$0318 & 19.50$\pm$0.06 & 19.32$\pm$0.05 & 18.78$\pm$0.06 & 18.14$\pm$0.09 & 19.79$\pm$0.10 & 19.64$\pm$0.14 & 19.53$\pm$0.13 & 19.63$\pm$0.17 & 19.39$\pm$0.25 & $>$19.0\\
J1104.3$+$0730 & 19.18$\pm$0.04 & 18.70$\pm$0.03 & 18.28$\pm$0.04 & 18.08$\pm$0.07 & 19.99$\pm$0.17 & 19.74$\pm$0.16 & 19.85$\pm$0.19 & \nodata & \nodata & \nodata\\
J1110.4$-$1835 & 18.91$\pm$0.08 & 18.41$\pm$0.04 & 18.52$\pm$0.06 & 17.97$\pm$0.10 & 20.73$\pm$0.15 & 19.86$\pm$0.13 & 19.64$\pm$0.12 & 19.14$\pm$0.12 & 18.82$\pm$0.15 & 18.98$\pm$0.33\\
J1224.6$-$8312 & 22.76$\pm$1.08 & 21.26$\pm$0.23 & 20.23$\pm$0.13 & 19.59$\pm$0.21 & $>$23.3 & $>$22.9 & $>$22.9 & $>$22.7 & $>$22.1 & $>$21.5\\
J1304.3$-$5535 & 18.76$\pm$0.10 & 18.40$\pm$0.06 & 18.10$\pm$0.06 & 17.72$\pm$0.06 & $>$19.7 & $>$19.6 & $>$19.6 & $>$19.7 & 18.82$\pm$0.23 & 18.57$\pm$0.29\\
J1326.6$-$5256 & 16.01$\pm$0.01 & 15.43$\pm$0.01 & 14.65$\pm$0.01 & 14.66$\pm$0.01 & 15.72$\pm$0.20 & 16.11$\pm$0.17 & 16.79$\pm$0.21 & 17.03$\pm$0.27 & $>$16.3 & 17.67$\pm$0.36\\
J1353.5$-$6640 & 17.66$\pm$0.14 & 17.18$\pm$0.06 & 16.83$\pm$0.06 & 16.31$\pm$0.13 & $>$19.0 & $>$19.2 & 18.62$\pm$0.27 & 18.22$\pm$0.14 & 17.74$\pm$0.13 & 17.69$\pm$0.17\\
J1427.6$-$3305 & 16.7$\pm$0.01 & 16.26$\pm$0.01 & 15.88$\pm$0.01 & 15.43$\pm$0.01 & 18.09$\pm$0.15 & 17.72$\pm$0.15 & 17.44$\pm$0.16 & 17.29$\pm$0.21 & 16.54$\pm$0.21 & $>$16.3\\
J1440.4$-$3845 & 17.33$\pm$0.04 & 17.02$\pm$0.03 & 16.65$\pm$0.07 & 16.40$\pm$0.14 & 17.76$\pm$0.07 & 17.69$\pm$0.09 & 17.55$\pm$0.09 & 17.45$\pm$0.09 & 17.05$\pm$0.10 & 16.83$\pm$0.16\\
J1508.7$-$4956 & 19.52$\pm$0.06 & 17.38$\pm$0.03 & 18.64$\pm$0.05 & 18.39$\pm$0.09 & $>$19.7 & $>$19.4 & $>$19.5 & $>$19.4 & 18.62$\pm$0.27 & $>$18.2\\
J1509.9$-$2951 & 20.07$\pm$0.18 & 19.83$\pm$0.12 & 19.93$\pm$0.17 & 19.10$\pm$0.22 & $>$21.3 & $>$20.9 & $>$21.0 & $>$20.7 & $>$20.1 & $>$19.5\\
J1525.2$-$5905 & \nodata & 15.88$\pm$0.19 & 16.13$\pm$0.15 & 15.71$\pm$0.10 & \nodata & \nodata & \nodata & \nodata & \nodata & \nodata\\
J1645.2$-$5747 & 15.03$\pm$0.07 & 14.63$\pm$0.04 & 14.59$\pm$0.05 & 14.75$\pm$0.11 & 18.71$\pm$0.29 & $>$18.8 & 17.69$\pm$0.18 & 18.47$\pm$0.16 & 15.10$\pm$0.05 & 16.27$\pm$0.07\\
J1745.4$-$0754 & 20.76$\pm$3.56 & 18.44$\pm$0.19 & 17.97$\pm$0.14 & 17.64$\pm$0.27 & $>$20.2 & $>$20.6 & $>$20.7 & $>$21.1 & $>$21.3 & $>$21.4\\
J2200.2$+$2139 & 17.56$\pm$0.01 & 17.07$\pm$0.01 & 16.83$\pm$0.01 & 16.93$\pm$0.01 & 19.35$\pm$0.17 & 19.08$\pm$0.19 & 18.58$\pm$0.16 & 18.39$\pm$0.19 & 17.64$\pm$0.19 & 17.22$\pm$0.28\\
J2227.8$+$0040 & 18.97$\pm$0.07 & 18.25$\pm$0.03 & 17.82$\pm$0.04 & 17.80$\pm$0.08 & 21.05$\pm$0.29 & 20.45$\pm$0.28 & 20.06$\pm$0.25 & $>$18.6 & 19.10$\pm$0.28 & 19.39$\pm$0.21\\
J2247.8$+$4413 & 17.50$\pm$0.01 & 16.80$\pm$0.01 & 16.69$\pm$0.01 & 16.90$\pm$0.01 & 18.55$\pm$0.05 & 18.94$\pm$0.12 & 18.24$\pm$0.09 & 17.99$\pm$0.08 & 17.62$\pm$0.09 & 17.42$\pm$0.15\\
J2307.4$-$1208 & 19.42$\pm$0.01 & 19.12$\pm$0.01 & 18.88$\pm$0.02 & 18.93$\pm$0.04 & 19.96$\pm$0.09 & 19.97$\pm$0.13 & 19.85$\pm$0.16 & 19.88$\pm$0.21 & 19.46$\pm$0.26 & $>$19.1\\
J2319.2$-$4207 & 16.16$\pm$0.01 & 15.05$\pm$0.01 & 14.59$\pm$0.01 & 14.60$\pm$0.01 & 18.96$\pm$0.08 & 18.99$\pm$0.12 & 18.35$\pm$0.09 & 15.42$\pm$0.05 & 16.36$\pm$0.05 & 17.66$\pm$0.07\\
J2324.7$+$0801 & 17.96$\pm$0.04 & 17.29$\pm$0.02 & 17.02$\pm$0.02 & 16.87$\pm$0.04 & 19.10$\pm$0.10 & 19.20$\pm$0.14 & 18.90$\pm$0.10 & 17.90$\pm$0.20 & 18.08$\pm$0.12 & 18.43$\pm$0.10\\
J2357.4$-$1716 & 17.80$\pm$0.01 & 17.29$\pm$0.01 & 16.99$\pm$0.01 & 17.19$\pm$0.03 & 18.28$\pm$0.07 & 18.27$\pm$0.10 & 18.20$\pm$0.08 & 17.82$\pm$0.18 & 17.88$\pm$0.10 & 18.02$\pm$0.02\\
\enddata
\end{deluxetable*}
\end{longrotatetable}
\end{center}


\startlongtable
\centering
\begin{deluxetable*}{ll|ccll|lclr}
	\tablecolumns{10}
	\tabletypesize{\small}
	\tablewidth{0pt}
	\tablecaption{\label{tab:result}SED fitting}
	\setlength{\tabcolsep}{0.05in} 
	\tablehead{
		\colhead{3FGL Name}  & \colhead{$z_{\rm phot, best}^a$}   & \multicolumn{4}{c}{Power Law Template} & \multicolumn{4}{c}{Galaxy Template} \\
		\cmidrule[\heavyrulewidth](lr){3-6}   \cmidrule[\heavyrulewidth](lr){7-10}
		\colhead{} & \colhead{} & \colhead{$z_{\rm phot}  ^b$} & \colhead{$\chi^2$} & \colhead{P$_{\rm z}^c$} & \colhead{$\beta^d$} \hspace{0.4cm} & \colhead{$z_{\rm phot}  ^b$} & \colhead{$\chi^2$} & \colhead{P$_{\rm z}^c$} & \colhead{Model}} 
	\startdata
	\multicolumn{10}{c}{\textbf{\small Sources with confirmed photometric redshifts}} \\
	\hline
J0427.3$-$3900 & ${1.42}^{+0.12}_{-0.10}$ & ${1.42}^{+0.12}_{-0.10}$ & 2.7 &  95.0 &  0.50 & ${1.25}^{+0.10}_{-0.10}$ & 2.1  & 89.6  & pl\_I22491\_20\_TQSO1\_80.sed  \\ 
J0609.4$-$0248 & ${1.73}^{+0.11}_{-0.10}$  & ${1.73}^{+0.11}_{-0.10}$ & 10.2 &  95.1 &  0.60 & ${1.34}^{+0.09}_{-0.06}$ & 7.5  & 99.3  & pl\_QSOH\_template\_norm.sed  \\ 
J1110.4$-$1835 & ${1.56}^{+0.09}_{-0.11}$ & ${1.56}^{+0.09}_{-0.11}$ & 9.0 &  95.8 &  0.90 &  ${1.42}^{+0.05}_{-0.05}$ & 14.2  & 97.4  & pl\_QSOH\_template\_norm.sed  \\
	\hline \hline\multicolumn{10}{c}{\textbf{\small Sources with photometric redshift upper limits}} \\
	\hline
J0009.1$+$0630 & $<$1.53 & ${1.38}^{+0.15}_{-0.33}$ & 11.0 & 41.0 & 1.35 & ${0.15 }^{+0.02 }_{-0.10 } $ & 8.3 & 83.4 & CB1\_0\_LOIII4.sed\\
J0013.2$-$3954 & $<$1.29 & ${1.11}^{+0.18}_{-0.00}$ & 17.1 & 32.7 & 1.25 & ${0.16}^{+0.06}_{-0.07} $ & 21.5 & 79.5 & I22491\_90\_TQSO1\_10.sed\\
J0019.4$+$2021 & $<$4.00 & ${3.93}^{+0.07}_{-0.21}$ & 3.6 & 88.7 & 2.45 & ${0.17}^{+0.04}_{-0.04} $ & 24.4 & 93.5 & S0\_90\_QSO2\_10.sed\\
J0026.7$-$4603 & $<$0.96 & ${0.23}^{+0.43}_{-0.00}$ & 14.8 & 24.8 & 0.65 & ${0.19}^{+0.00}_{-0.00} $ & 9.17 & 27.2 & pl\_TQSO1\_template\_norm.sed\\
J0041.9$+$3639 & $<$0.83 & ${0.03}^{+0.80}_{-0.00}$ & 10.1 & 16.5 &  0.90 & ${0.01}^{+0.00}_{-0.00} $ & 6.9 & 58.2 & pl\_I22491\_20\_TQSO1\_80.sed\\
J0116.3$-$6153 & $<$1.23 & ${0.24}^{+0.99}_{-0.00}$ & 5.2 & 24.1 & 1.20 & ${0.23}^{+0.09}_{-0.10} $  & 2.56 & 79.4 & I22491\_80\_TQSO1\_20.sed\\
J0125.4$-$2548 & $<$4.00 & ${4.00}^{+0.00}_{-0.20}$ & 6.1 & 100.0 & 2.00 & ${0.60}^{+0.12}_{-0.13} $ & 3.6 & 37.7 & S0\_30\_QSO2\_70.sed\\
J0133.0$-$4413 & \nodata & ${0.99}^{+0.01}_{-0.04}$ & 126.6 & 57.5 & 0.85 & ${0.57}^{+0.00}_{-0.01} $ & 20.2 & 100.0 & M82\_template\_norm.sed\\
J0143.7$-$5845 & $<$1.18 & ${0.24}^{+0.94}_{-0.00}$ & 16.9 & 25.8 & 0.30 & ${0.25}^{+0.02}_{-0.00} $ & 14.3 & 58.7 & pl\_QSO\_DR2\_029\_t0.spec\\
J0253.0$-$0125 & \nodata & ${0.87}^{+0.02}_{-0.00}$ & 100.6 & 27.5 & 0.40 & ${0.59}^{+0.02}_{-0.03} $ & 61.0 & 63.0 & CB1\_0\_LOIII4.sed\\
J0304.3$-$2836 & \nodata & ${0.93}^{+0.01}_{-0.09}$ & 154.4 & 39.0 & 0.75 & ${0.64}^{+0.00}_{-0.01} $ & 51.6 & 100.0 & CB1\_0\_LOIII4.sed\\
J0316.2$-$6436 & $<$1.10 & ${0.24}^{+0.86}_{-0.00}$ & 17.2 & 28.0 & 0.65 & ${0.03}^{+0.00}_{-0.00} $ & 16.9 & 73.4 & pl\_QSO\_DR2\_029\_t0.spec\\
J0335.3$-$4459 & \nodata & ${0.24}^{+0.49}_{-0.00}$ & 84.0 & 26.0 & 1.20 & ${0.13}^{+0.11}_{-0.03} $ & 79.6 & 75.3 & I22491\_70\_TQSO1\_30.sed\\
J0359.3$-$2612 & $<$3.08 & ${2.42}^{+0.66}_{-1.37}$ & 0.79 & 27.2 & 1.15 & ${0.06}^{+0.05}_{-0.00} $ & 0.31 & 12.3 & CB1\_0\_LOIII4.sed\\
J0409.8$-$0358 & \nodata & ${0.45}^{+0.64}_{-0.00}$ & 75.6 & 26.1 & 1.40 & ${0.02}^{+0.00}_{-0.00} $ & 8.8 & 87.8 & I22491\_80\_TQSO1\_20.sed\\
J0529.2$-$5917 & $<$1.55 & ${1.37}^{+0.18}_{-0.16}$ & 6.6 & 61.9 & 1.55 & ${1.14}^{+0.10}_{-0.09} $ & 8.5 & 96.4 & Spi4\_template\_norm.sed\\
J0556.0$-$4353 & $<$1.25 & ${1.07}^{+0.18}_{-0.00}$ & 2.9 & 31.4 & 1.20 & ${0.03}^{+0.00}_{-0.00} $ & 5.2 & 73.4 & I22491\_70\_TQSO1\_30.sed\\
J0627.9$-$1517 & \nodata & \nodata & \nodata & \nodata & \nodata & ${0.72}^{+0.00}_{-0.00} $ & 173.1 & 82.5 & S0\_template\_norm.sed\\
J0807.1$-$0541 & $<$1.24 & ${0.96}^{+0.28}_{-0.00}$ & 15.3 & 32.3 & 1.35 & ${0.01}^{+0.00}_{-0.00} $ & 15.6 & 99.6 & I22491\_80\_TQSO1\_20.sed\\
J0958.3$-$0318 & \nodata & ${0.12}^{+0.83}_{-0.00}$ & 27.1 & 23.0 & 0.60 & ${0.73}^{+0.06}_{-0.05} $ & 8.4 & 91.3 & I22491\_70\_TQSO1\_30.sed\\
J1104.3$+$0730 & \nodata & \nodata & \nodata & \nodata & \nodata & \nodata & \nodata & \nodata & \nodata\\
J1224.6$-$8312 & $<$4.00 & ${4.00}^{+0.00}_{-0.10}$ & 3.1 & 98.5 &  2.0 & ${0.73}^{+0.70}_{-0.28} $ & 0.01 & 26.2 & S0\_30\_QSO2\_70.sed\\
J1427.6$-$3305 & $<$1.15 & ${0.48}^{+0.67}_{-0.00}$ & 5.5 & 25.3 & 1.55 & ${0.01}^{+0.00}_{-0.00} $ & 11.8 & 78.0 & I22491\_90\_TQSO1\_10.sed\\
J1440.4$-$3845 & $<$0.98 & ${0.12}^{+0.86}_{-0.00}$ & 9.4 & 22.2 & 0.70 & ${0.15}^{+0.28}_{-0.06} $ & 6.2 & 37.9 & pl\_I22491\_20\_TQSO1\_80.sed\\
J1509.9$-$2951 & $<$3.45 & ${2.97}^{+0.48}_{-0.59}$ & 1.9 & 40.7 & 0.30 & ${2.17}^{+0.50}_{-0.00} $ & 0.6 & 43.7 & M82\_template\_norm.sed\\
J1645.2$-$5747 & \nodata & ${1.87}^{+0.11}_{-0.04}$ & 417.7 & 98.4 & 3.00 & ${0.89}^{+0.11}_{-0.01} $ & 390.1 & 95.2 & S0\_template\_norm.sed\\
J2200.2$+$2139 & $<$1.64 & ${1.53}^{+0.11}_{-0.12}$ & 14.1 & 71.2 & 1.20 & ${0.08}^{+0.05}_{-0.06} $ & 11.6 & 97.8 & Spi4\_template\_norm.sed\\
J2227.8$+$0040 & $<$1.55 & ${1.37}^{+0.18}_{-0.41}$ & 5.2 & 51.4 & 1.75 & ${0.29}^{+0.02}_{-0.02} $ & 23.0 & 67.5 & Mrk231\_template\_norm.sed\\
J2247.8$+$4413 & \nodata & ${0.82}^{+0.22}_{-0.00}$ & 31.2 & 32.4 & 1.25 & ${0.01}^{+0.00}_{-0.00} $ & 28.2 & 92.6 & I22491\_70\_TQSO1\_30.sed\\
J2307.4$-$1208 & $<$0.99 & ${0.99}^{+0.00}_{-0.00}$ & 4.0 & 9.8 & 0.75 & ${0.01}^{+0.57}_{-0.00} $ & 3.1 & 22.4 & pl\_I22491\_10\_TQSO1\_90.sed\\
J2319.2$-$4207 & \nodata & ${3.93}^{+0.01}_{-0.01}$ & 172.0 & 99.9 & 3.00 & ${0.08}^{+0.00}_{-0.02} $ & 16.1 & 100.0 & Sey2\_template\_norm.sed\\
J2324.7$+$0801 & $<$0.97 & ${0.03}^{+0.94}_{-0.00}$ & 8.4 & 14.5 & 1.50 & ${0.02}^{+0.00}_{-0.00} $ & 11.6 & 64.6 & I22491\_80\_TQSO1\_20.sed\\
J2357.4$-$1716 & $<$0.89 & ${0.23}^{+0.66}_{-0.00}$ & 15.2 & 26.6 & 0.80 & ${0.23}^{+0.00}_{-0.00} $ & 8.7 & 20.2 & pl\_I22491\_20\_TQSO1\_80.sed\\
    \enddata
	\tablenotetext{a}{Best photometric redshift.}
	\tablenotetext{b}{Uncertainties on photometric redshifts and the upper limits are reported at 1$\sigma$ confidence level}
	\tablenotetext{c}{Redshift probability density at $z_{phot} \pm 0.1(1+z_{phot})$}
	\tablenotetext{d}{Spectral slope for power law model of the form $F_{\lambda} \propto \lambda^{-\beta}$}
\end{deluxetable*}

\bibliographystyle{apj}
\bibliography{Swift_Sara_ms_ma}
\end{document}